\documentclass[a4paper,twocolumn,english,aps,prl,superscriptaddress]{revtex4-1}
\usepackage[LGR,T1]{fontenc}
\usepackage[latin9]{inputenc}
\usepackage{babel}
\usepackage{float}
\usepackage{booktabs}
\usepackage{amsmath}
\usepackage{amssymb}
\usepackage{graphicx}
\usepackage[unicode=true,pdfusetitle,
 bookmarks=true,bookmarksnumbered=false,bookmarksopen=false,
 breaklinks=false,pdfborder={0 0 1},backref=false,colorlinks=false]
 {hyperref}

\makeatletter

\DeclareRobustCommand{\greektext}{%
  \fontencoding{LGR}\selectfont\def\encodingdefault{LGR}}
\DeclareRobustCommand{\textgreek}[1]{\leavevmode{\greektext #1}}
\ProvideTextCommand{\~}{LGR}[1]{\char126#1}

\providecommand{\tabularnewline}{\\}

\usepackage{babel}
\usepackage{babel}

\makeatother

\begin{document}
\title{Topological States in Generalized Electric Quadrupole Insulators}
\author{Chang-An Li }
\email{changan@connect.hku.hk}
\affiliation{School of Science, Westlake University, 18 Shilongshan Road, Hangzhou
310024, Zhejiang, China}
\affiliation{Institute of Natural Sciences, Westlake Institute for Advanced Study,
18 Shilongshan Road, Hangzhou 310024, Zhejiang, China}
\author{Shu-Shan Wu}
\affiliation{Department of Physics, Hangzhou Normal University, Hangzhou 310036,
Zhejiang, China }
\date{\today }
\begin{abstract}
The modern theory of electric polarization has recently been extended
to higher multipole moments, such as quadrupole and octupole moments. The higher electric multipole insulators are essentially
topological crystalline phases protected by underlying crystalline symmetries. Henceforth, it is natural
to ask what are the consequences of symmetry breaking in these higher multipole insulators.
In this work, we investigate topological phases and the consequences of
symmetry breaking in generalized electric quadrupole insulators. Explicitly,
we generalize the Benalcazar-Bernevig-Hughes model by adding specific
terms in order to break the crystalline and non-spatial
symmetries. Our results show that chiral symmetry breaking induces an indirect
gap phase which hides corner modes in bulk bands, ruining the topological
quadrupole phase. We also demonstrate that quadrupole moments can remain quantized
even when mirror symmetries are absent in a generalized model.  Furthermore,
it is shown that topological quadrupole phase is robust against a
unique type of disorder presented in the system.
\end{abstract}
\maketitle

\section{Introduction}

Electric polarization in crystalline had traditionally been a long-standing issue
as it is not a well-defined observable that can simply be given
by the expectation value of a local operator \cite{Resta94rmp,Resta98prl}.
The modern theory of bulk polarization is based on the Berry phase, which
is determined via the wave functions of energy bands over a closed path
in the Brillouin zone \cite{King93prb}. The theory has exerted a strong influence
on condensed matter physics over recent decades, particularly
on the development of topological band insulators \cite{XiaoD10rmp,Kane10rmp,QiXL11rmp,SQS,BernevigBook,Thouless83prb,FuL06prb}.
Recent studies have extended the modern theory of polarization to higher
multipole moments, such as quadrupole and octupole moments. As is
well known, quantized bulk polarization in the Su-Schrieffer-Heeger (SSH)
model gives rise to the fractional charge $\pm e/2$ at the boundaries
of a one-dimensional (1D) sample, where $e$ is the electron charge \cite{SSH79prl}.
Benalcazar \textit{et al}. extended this to two- and
three-dimensional systems that hold quantized bulk quadrupole and
octupole moments, respectively \cite{Benalcazar17Science,BBH17prb}.
These systems are denoted as electric quadrupole and octupole
insulators. Explicit models demonstrate that these quantized bulk
multipole moments also manifest fractionalized boundary charges. Traditionally,
a topological bulk state in a $d$-dimensional system has robust ($d-1$) dimensional
boundary states. Nevertheless, topological quadrupole (octupole) insulators
have localized states at corners, namely, the $d-2$ ($d-3$) boundaries
of the system. This ``high-order'' bulk-boundary correspondence
\cite{Trifunovic19prx} casts these topological phases into a new
class of topological insulators denoted as ``high-order topological insulators''
\cite{Schindler18SA}. Generally, a $d$-dimensional high-order topological
insulator has non-trivial boundary states at the $d-m$ boundary ($d\geq m\geq2$).
Thus, such high-order topological insulators have attracted
much theoretical and experimental interest over the past few years \cite{Langbehn17prl,Khalaf18prb,SongZD19prl,Geier18prb,petrides2019arxiv,WangZJ19prl,Schindler18NP,Serra-Garcia18nature,Peterson18nature,Franca18prb,Thomale18np},
and have been extended to high-order topological superconductors \cite{YanZB18prl,WangQ18prl,Hsu18prl,Volpez19prl,LiuT18prb,ZhuXY18prb,Shapourian18prb}
and even semi-metals \cite{Ezawa18prl,okugawa2019arxiv,wieder2019arxiv}.

Higher electric multipole insulators are essentially topological
crystalline insulators \cite{FuL11prl,Neupert18springer,Jan17prx},
with the quantization of multipole moments imposed via the underlying
crystalline symmetries of the system. For example, the quantization
of quadrupole (octupole) moments of the Benalcazar-Bernevig-Hughes (BBH) model
can be performed using a combination of mirror symmetries \cite{BBH17prb}.
As well as crystalline symmetries, high-order topological insulators
may also require non-spatial symmetries (i.e., chiral, time-reversal,
and particle-hole symmetries) in order to protect their high-order topology \cite{Schindler18SA,Yoshida18prb,Khalaf18prx}.
A key topic of interest is the determination of the consequences of symmetry breaking,
including both the crystalline and non-spatial symmetries,
in higher electric multipole insulators. In simple terms, crystalline
symmetry breaking damages topological phases while non-spatial symmetry breaking is irrelevant. However, the consequences are in fact more complicated. Furthermore, the high-order topology in
the BBH model is characterized by the so-called
``nested Wilson loop''. This characterization is based on the
equivalent topology between the Wannier bands and the edge spectrum
\cite{YuR11prb,Fidkowski11prl,Alexandradinata14prb,khalaf19arxiv}.
The equivalence, however, may be lost under certain circumstances
\cite{Yang19arxiv,Yang2019arxiv}.
In this case it is also meaningful
to ask how to demonstrate high-order phase if nested Wilson loop approach
fails.

In this paper, we exploit several generalized BBH
models in order to evaluate newly appearing topological phases as well as the consequences
of symmetry breaking in electric quadrupole insulators. In the first
generalized model, we include additional hopping terms to break chiral
symmetry. These chiral
symmetry breaking terms lead the system to an indirect gap phase.
Corner modes will be buried by bulk bands in the indirect gap phase,
and quadrupole moments are not well defined. The nested
Wilson loop approach fails to capture this phase since there is no
real topological phase transition. In the second model, we include hopping
terms with imaginary amplitudes in order to break the time-reversal and chiral
symmetries. Following an induced phase transition, the equivalent topology between the Wannier bands and edge spectrum may be lost in this simple model. In this case, the nested Wilson loop approach is no longer
applicable since its basis is ruined, while the quantized quadrupole
moments together with edge polarization and fractional corner charges
remain effective for the characterization of high-order topology. In the
third model, we focus on mirror symmetry breaking while keeping inversion
symmetry. Unexpectedly, the
quadrupole moments remain quantized, despite the breakdown of mirror symmetry. Note that the $C_{2}$ (inversion) symmetry is kept for all these three models, and it is necessary for the well-defined quadrupole moments.
More interestingly, we determine the quantized quadrupole moments to be
robust against disorders of a unique type added to the system.

The remainder of this paper is organized as follows. Sec. II introduces
the extended model with chiral symmetry breaking, and discusses the
consequences of the indirect gap phase. Sec. III presents the inclusion of
time-reversal and chiral symmetry breaking in the model. Sec. IV details the results
of mirror symmetry breaking. Sec. V considers the robustness
of quantized quadrupole moments in the presence of disorders. Finally,
Sec. VI concludes our results with a discussion.

\section{Indirect gap phases: chiral symmetry breaking}

\begin{figure}
\includegraphics[width=1\linewidth]{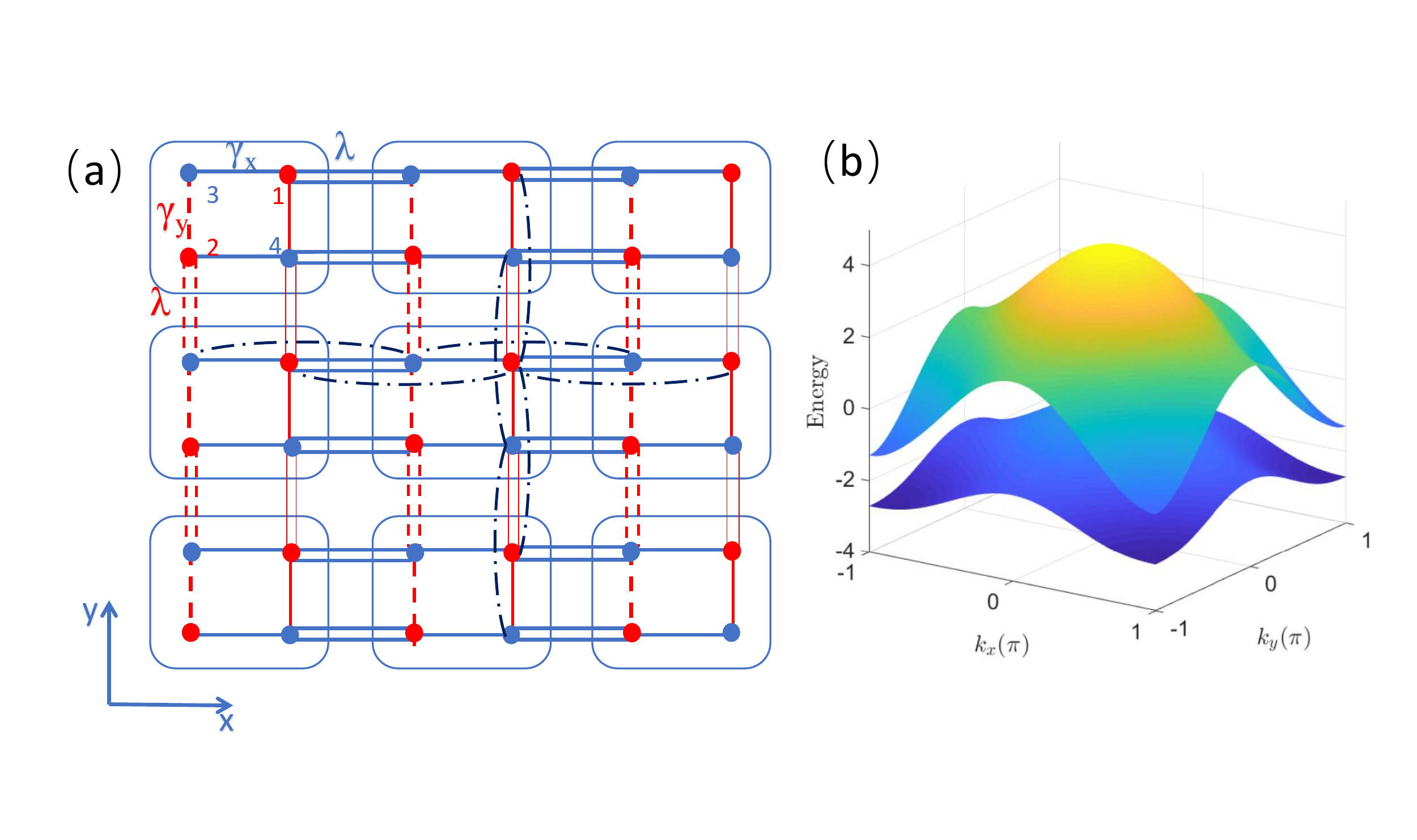}

\caption{(a) Lattice structure of extended BBH model $H_{1}$. The dashed dark
blue lines represent hopping at equivalent sites with strength $t$
between nearest unit cells. For simplicity, here we only sketch parts
of this kind of hopping. (b) Energy bands corresponding to (a) with
parameters fixed at $\gamma_{x}=\gamma_{y}=0.5,$ and $t=0.5$. Here
the system has an indirect gap. \label{fig:Extended1}}
\end{figure}

The topology of the SSH model, which forms the basis of the BBH model, is protected by
chiral symmetry. Thus, in this section we consider a generalized BBH
model with chiral symmetry breaking by introducing hopping terms between
equivalent sites, as presented in Fig. \ref{fig:Extended1}(a). The
dressed model is expressed as follows:

\begin{equation}
H_{1}=H_{0}+\sum_{\mathbf{R}}\text{\ensuremath{\sum_{\zeta=1}^{4}\sum_{s=x,y}}}t(C_{\mathbf{R},\zeta}^{\dagger}C_{\mathbf{R}+\hat{s},\zeta}+\mathrm{H.c.}),\label{eq:H1}
\end{equation}
where $t$ is the corresponding hopping amplitude, and
\begin{alignat}{1}
H_{0}= & \sum_{\mathbf{R}}\Big[\gamma_{x}(C_{\mathbf{R},1}^{\dagger}C_{\mathbf{R},3}+C_{\mathbf{R},2}^{\dagger}C_{\mathbf{R},4})\nonumber \\
 & +\gamma_{y}(C_{\mathbf{R},1}^{\dagger}C_{\mathbf{R},4}-C_{\mathbf{R},2}^{\dagger}C_{\mathbf{R},3})\nonumber \\
 & +\lambda(C_{\mathbf{R},1}^{\dagger}C_{\mathbf{R}+\hat{x},3}+C_{\mathbf{R},4}^{\dagger}C_{\mathbf{R}+\hat{x},2})\nonumber \\
 & +\lambda(C_{\mathbf{R},1}^{\dagger}C_{\mathbf{R}+\hat{y},4}-C_{\mathbf{R},3}^{\dagger}C_{\mathbf{R}+\hat{y},2})\Big]+\mathrm{H.c.},
\end{alignat}
The operators $C_{\mathbf{R},\zeta}^{\dagger}$ ($C_{\mathbf{R},\zeta}$)
are creation (annihilation) operators at unit cell $\mathbf{R}=(m\hat{x},n\hat{y})$
with $\zeta=1,2,3,4$ being orbital-like degree of freedoms. Here
the parameters $\gamma_{x,y}$ and $\lambda$ are hopping amplitudes
within and between unit cells, as sketched in  Fig. \ref{fig:Extended1}(a).
The Hamiltonian $H_{0}$ represents the original BBH model. Note the
lattice constant is assumed to be unity and $\lambda=1$ throughout
the following sections.

The corresponding Bloch Hamiltonian in momentum space is described as follows:
\begin{equation}
H_{1}({\bf k})=H_{q}({\bf k})+2t\sum_{s=x,y}\cos k_{s},\label{eq:H1_k}
\end{equation}
where
\begin{alignat}{1}
H_{q}({\bf k}) & =[\gamma_{x}+\lambda\cos k_{x}]\Gamma_{4}+\lambda\sin k_{x}\Gamma_{3}\nonumber \\
 & +[\gamma_{y}+\lambda\cos k_{y}]\Gamma_{2}+\lambda\sin k_{y}\Gamma_{1}.\label{eq:H_q}
\end{alignat}
The Gamma matrices are defined as $\Gamma_{j}=-\tau_{2}\sigma_{j}$
($j=1,2,3$), and $\Gamma_{4}=\tau_{1}\sigma_{0}$ with $\tau$ and
$\sigma$ both being Pauli matrices for the unit cell orbital degrees
of freedom. Due to the appearance of identity terms in Eq.~\eqref{eq:H1_k},
chiral symmetry is no longer preserved. One can check that

\begin{equation}
\mathcal{C}H_{1}({\bf k})\mathcal{C}^{-1}\neq-H_{1}({\bf k}),\ \mathcal{C}=\tau_{3}\sigma_{0},
\end{equation}
where $\mathcal{C}$ is the chiral symmetry operator. Particle-hole
symmetry is not preserved either, which is obvious from the energy
spectrum $E_{\pm}=2t\sum_{s=x,y}\cos k_{s}\pm\sqrt{\epsilon_{x}^{2}(k_{x})+\epsilon_{y}^{2}(k_{y})}$
where $\epsilon_{s}^{2}(k_{s})=\gamma_{s}^{2}+2\gamma_{s}\lambda\cos k_{s}+\lambda^{2}$.
Since the system respects both time-reversal symmetry and inversion symmetry, its
energy bands are still doubly degenerated. The global symmetries
cast the system into AI class, and its classification is trivial in
two-dimensional (2D) \cite{Schnyder08prb,ChiuCK16rmp}. Note that Eq.~\eqref{eq:H1_k}
keeps mirror symmetries.

\begin{figure}
\includegraphics[width=1\linewidth]{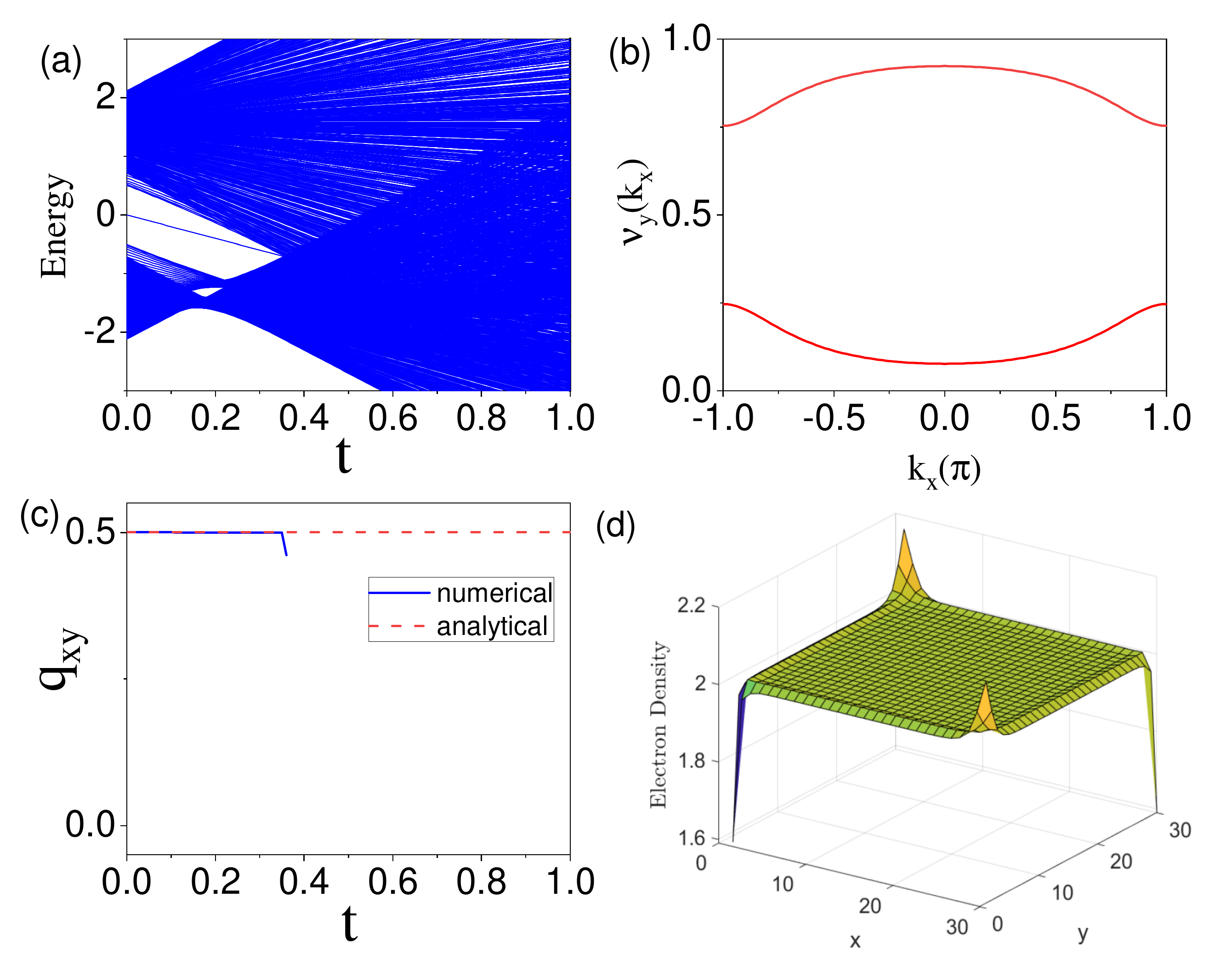}

\caption{Indirect gap phase induced by chiral symmetry breaking. (a) Energy
spectrum of $H_{1}$ as functions of parameter $t$. (b) Wannier bands for fixed
$t=0.5$. (c) Calculated quadrupole moments when varying parameter $t$ (red
line is calculated analytically with nested Wilson loop approach,
blue line is calculated numerically under periodic boundary condition
with size $L_{x}\times L_{y}=40\times40$). (d) Electron charge density
in the nontrivial phase with $t=0.2$. Other parameters are $\gamma_{x}=\gamma_{y}=0.5$.
\label{fig:Extended1_phase}}
\end{figure}

Due to the breakdown of chiral symmetry, the emerging ``new phase'' is an indirect gap phase
\cite{chen17arxiv,LiLH14prb,Prezgonzlez18arxiv}.
Figure \ref{fig:Extended1}(b) demonstrates that the
valence band $E_{\mathrm{max}}^{\mathrm{v}}$ maximum exceeds the
conduction band $E_{\mathrm{min}}^{\mathrm{c}}$ minimum. However, there
is no bulk gap closure. We denote this phase as an indirect gap phase.
The appearance of the indirect gap phase is determined
by the following inequality:

\begin{equation}
t>\frac{1}{8}\sum_{\alpha=\pm}\sqrt{\sum_{s=x,y}(\gamma_{s}+\alpha\lambda)^{2}}.\label{eq:indirect}
\end{equation}
The topological phase of an electric quadrupole insulator is characterized by quadrupole moments $q_{xy}$  \cite{Benalcazar17Science,BBH17prb}.  Classically, the quadrupole moments are defined as $q_{ij}=\int d^{3}r\rho(r)r_{i}r_{j}$ where $\rho(r)$ is the charge density and $r_{i,j}$ are positions with $i,j=x,y$. Benalcazar \textit{et al}. proposed insulators having only nonvanishing $q_{xy}$ which is quantized by mirror symmetries.  Quadrupole moments will
manifest hierarchical observable properties such as edge polarizations (see Appendix B.3) and corner charges (see Appendix B.4). Note that the indirect gap phase here can hide the corner modes in real space. Despite the varying of parameter $t$, the appearance of indirect gap phase is not reflected by the Wannier sector polarization ${\bf p}^{\nu}$ (see Appendix B.2)
or the quadrupole moments $q_{xy}$ calculated using nested Wilson loop (see Appendix B.5).
In addition, Fig. \ref{fig:Extended1_phase}(b) shows
that the Wannier bands are not affected by the increase of parameter $t$ at all.
This is a reasonable observation since no real gap closure process occurs,
and the eigen states are unchanged as the second
term proportional to $t$ in Eq.~\eqref{eq:H1_k} is an identity
term. Thus, Eq.~\eqref{eq:Wilsonloop} in Appendix
B.1 indicates that the Wannier bands will be independent of parameter $t$.

\begin{figure}
\includegraphics[width=1\linewidth]{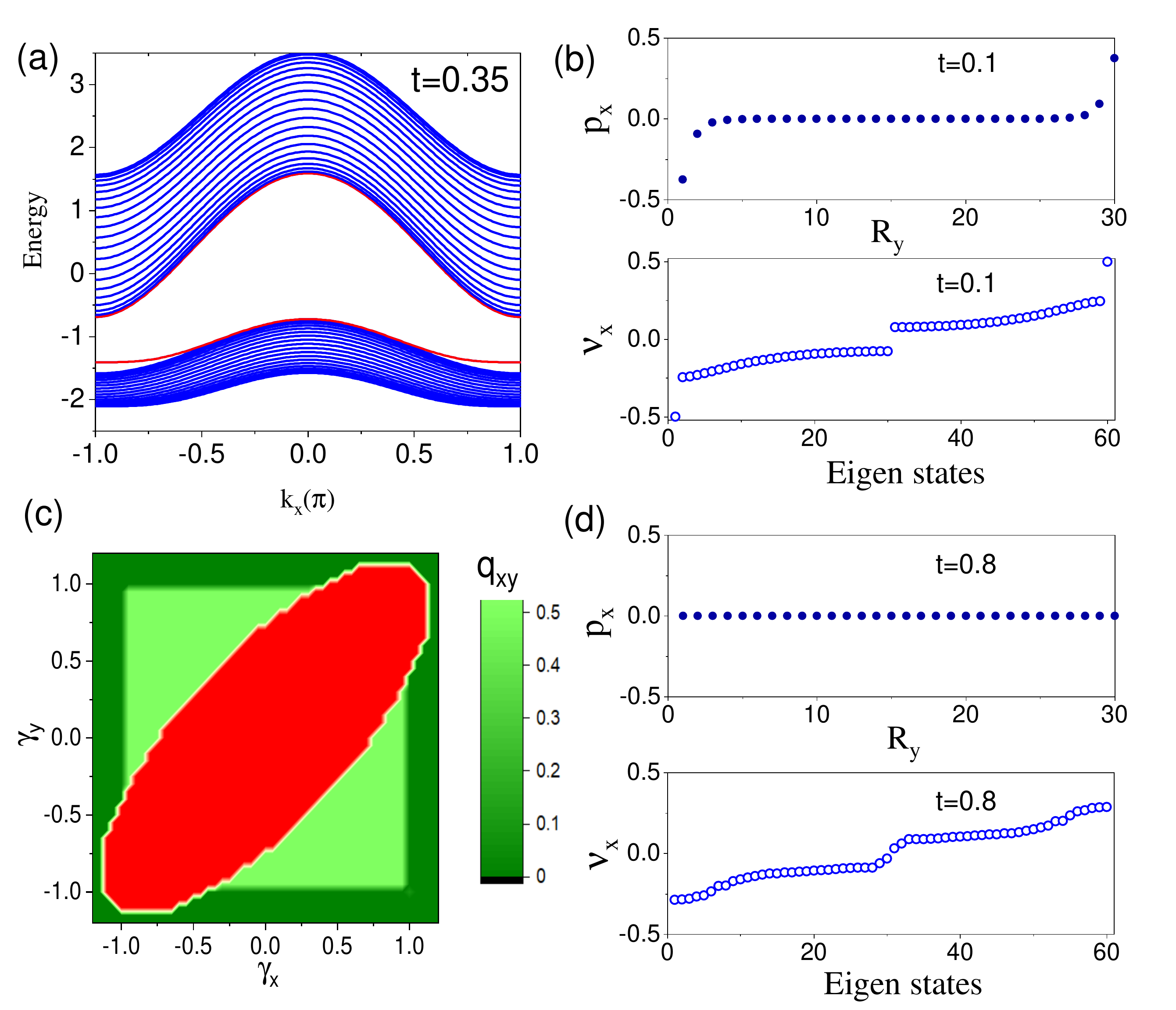}

\caption{Signatures of the indirect gap phase of Hamiltonian $H_{1}$. (a) Energy
spectrum of $H_{1}$ at a ribbon with width $L_{y}=20$ for $t=0.35$.
Red lines denote edge states. Panels (b) and (d) plot the edge polarization
and Wannier centers of states for $t=0.1$ and $0.8$, respectively.
(c) Phase diagram modified by the chiral breaking term in $H_{1}$
with fixed parameter $t=0.38$. Here $q_{xy}$ is calculated numerically under
periodic boundary condition with size $L_{x}\times L_{y}=30\times30$.
Red color region denotes indirect gap phases. Other parameters are $\gamma_{x}=\gamma_{y}=0.5$. \label{fig:Extended1_phase2}}
\end{figure}

Alternatively, we also employ a real-space recipe to calculate the
electric quadrupole moment \cite{Kang19prb,Wheeler19prb}, which is numerically
calculated as follows:

\begin{equation}
q_{xy}=\frac{1}{2\pi}\mathrm{Imlog}\langle\Psi_{G}|\hat{U}_{2}|\Psi_{G}\rangle,
\end{equation}
where $|\Psi_{G}\rangle$ is the many-body ground states, and $\hat{U}_{2}\equiv\exp[i2\pi\hat{q}_{xy}]$
with $\hat{q}_{xy}=\sum_{{\bf R}}xy\hat{n}({\bf R})/(L_{x}L_{y})$
as quadrupole momentum density operator per unit cell at position
${\bf R}$. Here $L_{s=x,y}$ are the length of sample along $\hat{s}$
direction. Note that we need to eliminate the atomic
limit contribution \cite{roy19arxiv}. Despite its defects \cite{Ono19prb}, this technique proves to be effective.

We now further evaluate Fig. \ref{fig:Extended1_phase}. Prior to
entering the indirect gap phase, the quantized quadrupole moments
$q_{xy}=1/2$ remain, accompanied by sharply localized corner states
(see Appendix B.4). At the same time, the corresponding energy of the corner states
is shifted away from zero-energy. Figure \ref{fig:Extended1_phase}(d) shows that
the charge density is not ``symmetric'' with respect to different
corners, yet the integrated charge over a quarter of the sample around
each corner still quantizes to $\pm e/2$. After the ``transition''
point, the corner states are hidden by bulk valence bands, and the
quantized quadrupole moments are destroyed. The buried corner states
work similarly as hidden helical edge states in topological insulators \cite{LiC18prb,Skolasinski18prb}.
In this case, there are no gaps in the system, thus the real-space recipe is not
applicable.

It is evident that the original non-trivial phases (see Appendix B.6)
will shrink due to the appearance of indirect gap phases. As demonstrated in the
energy spectrum in Fig. \ref{fig:Extended1_phase2}(a), the conduction
bands ``overlap'' with the valence bands, and the edge states merge to
bulk bands. The phase diagram Fig. \ref{fig:Extended1_phase2}(c) presents
a closed region that represents the indirect gap phase, which
is consistent with the restriction condition of Eq.~\eqref{eq:indirect}.
Furthermore, both the edge polarization $p_{x}(R_{y})$ (see Appendix B.3) and the
Wannier values of states $\nu_{x}$ (see Appendix B.1) can
aid in the detection of topological phases. Note that $p_{x}$ and
$\nu_{x}$ are calculated on a ribbon with a finite width. Prior to entering
the indirect gap phase, $p_{x}(R_{y})$ exhibits non-zero values, with an integrated value
over half of the ribbon width as $\pm e/2$.
Moreover, there are two isolated topological modes associated with the Wannier center
$\nu_{x}=\pm1/2$. In contrast, for the indirect gap phase with larger $t$,
the edge polarization vanishes, and the quantized Wannier value states
disappear. This is because the edge states totally merge
into the bulk bands.

\section{Discrepant topology between Wannier bands and edge spectrum}

\begin{figure}
\includegraphics[width=1\linewidth]{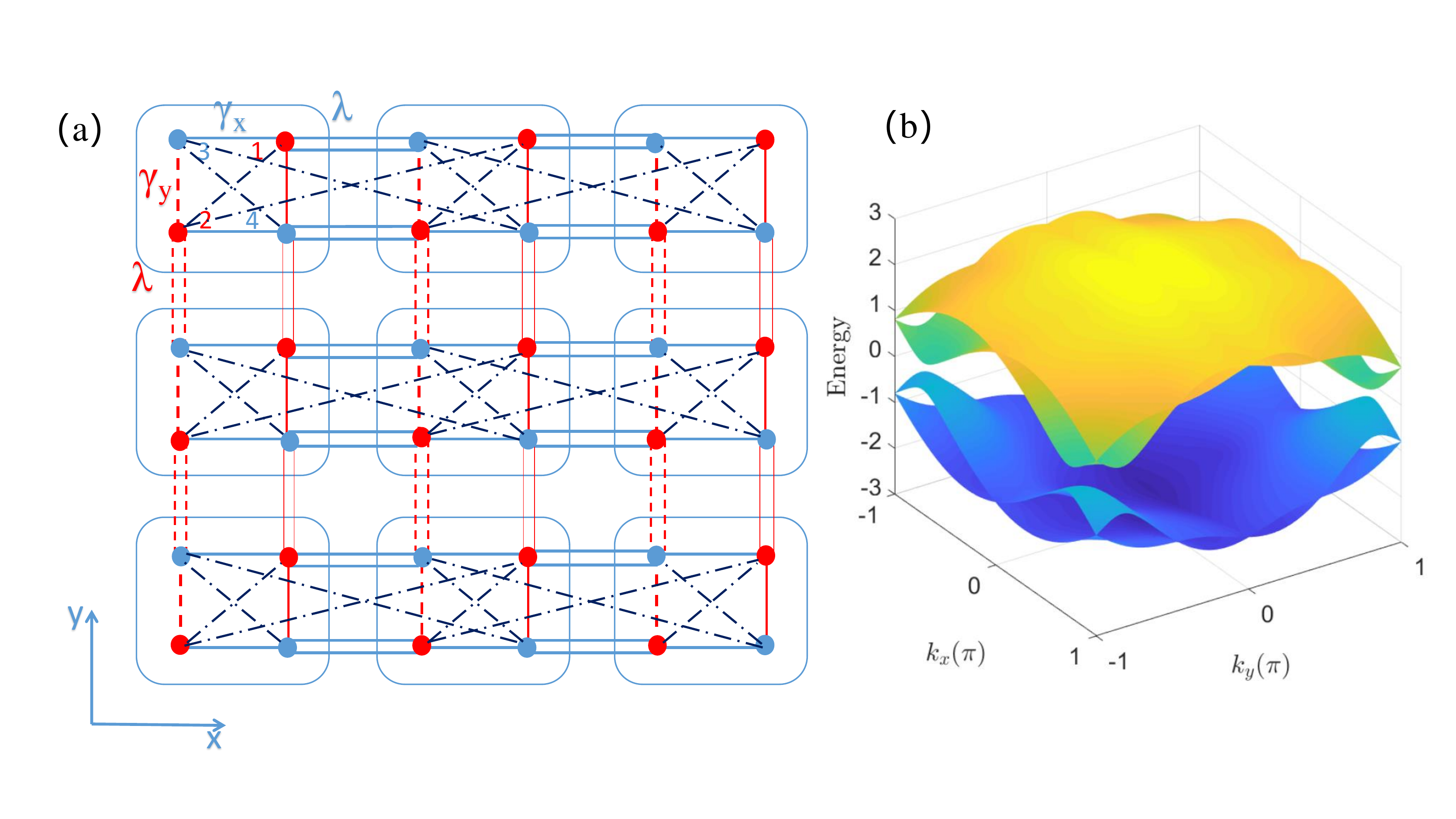}

\caption{(a) Lattice structure of extended BBH model $H_{2}$. The dashed dark
blue lines represent hopping within and between unit cells. These
hopping with imaginary amplitudes will break time-reversal symmetry.
(b) Energy bands corresponding to (a) with parameters fixed at $\gamma_{x}=0.8,\gamma_{y}=0.5,$
and $J_{2}=0.3.$ The band degeneracy is lifted. \label{fig:Extended2}}
\end{figure}

In this section, we consider the consequences of non-spatial symmetry
breaking (time-reversal and chiral symmetry) in the second generalized
BBH model. The imaginary hopping terms are introduced as depicted in
Fig. \ref{fig:Extended2}(a). The corresponding modified Hamiltonian
is described as follows:

\begin{alignat}{1}
H_{2} & =H_{0}+\sum_{\mathbf{R}}J_{2}(iC_{\mathbf{R},2}^{\dagger}C_{\mathbf{R},1}-iC_{\mathbf{R},4}^{\dagger}C_{\mathbf{R},3}+\mathrm{H.c.})\nonumber \\
 & +\sum_{\mathbf{R}}J_{2}(-iC_{\mathbf{R},2}^{\dagger}C_{\mathbf{R}+\hat{x},1}-iC_{\mathbf{R},3}^{\dagger}C_{\mathbf{R}+\hat{x},4}+\mathrm{H.c.}),\label{eq:H2}
\end{alignat}
where $iJ_{2}$ is the imaginary hopping amplitude ($J_{2}$ is a real
number). It is imaginary due to the enclosed half-$\pi$ flux, with the
sign convention as in Eq.~\eqref{eq:H2}. In the original BBH model, there is $\pi$ flux threading each square plaquette to open band gap, thus for our extended model, we can choose a gauge such that the triangle, for example, triangle with points  $1\rightarrow 2\rightarrow4\rightarrow1$ in Fig. \ref{fig:Extended2}, contains half-$\pi$ flux. The hopping terms with
imaginary amplitudes break the time-reversal symmetry \cite{Haldane88prl}.
For simplicity, only $iJ_{2}$ hopping along the $x$ direction is considered,
and we assume the same amplitudes within and between unit cells.

\begin{figure}
\includegraphics[width=1\linewidth]{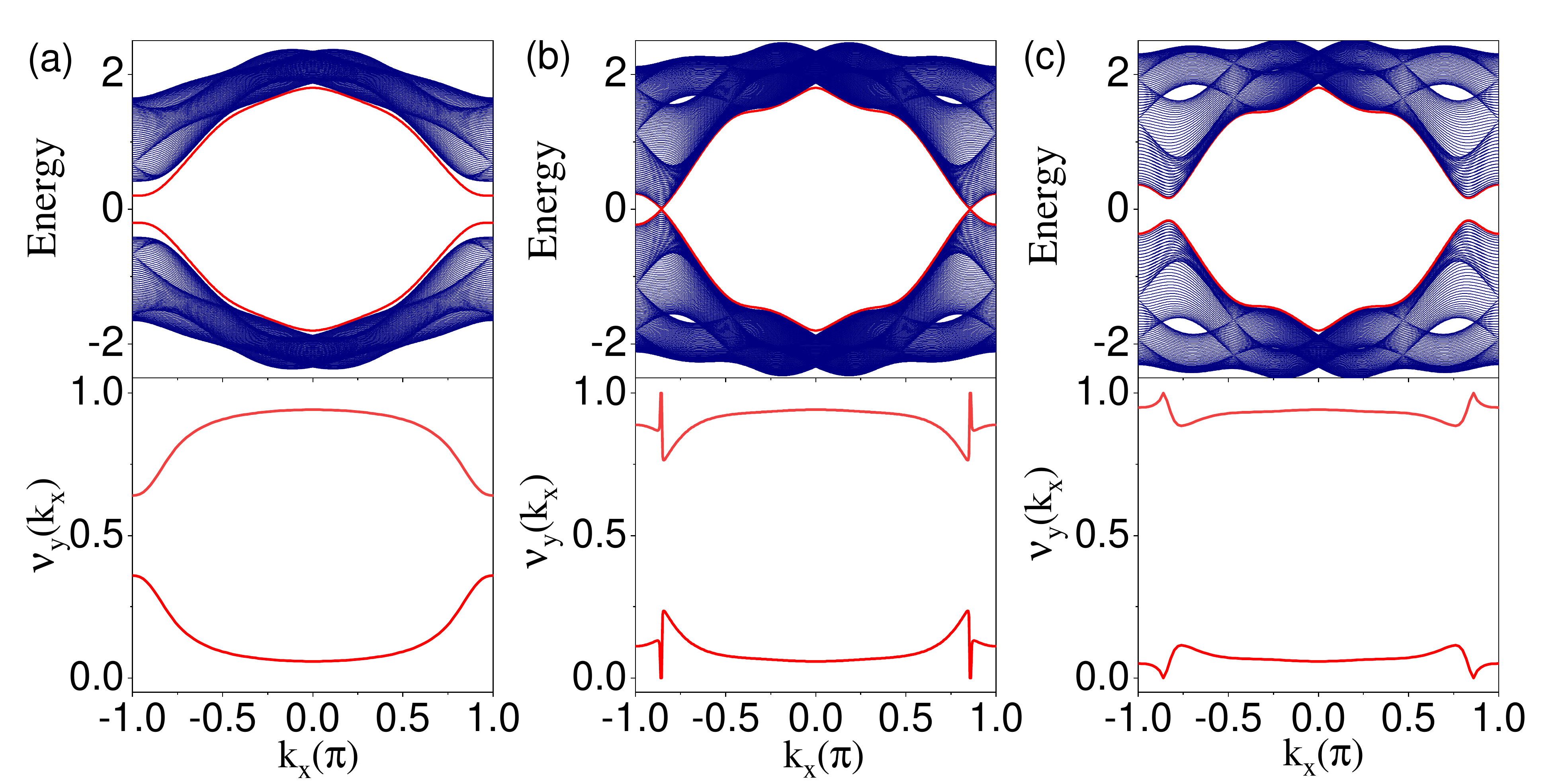}

\caption{Edge spectrum and corresponding Wannier bands for extended model $H_{2}$.
(a) $J_{2}=0.2$. (b) $J_{2}=0.5$. Gap closure happens both at edge
spectrum and Wannier bands. (c) $J_{2}=0.6$. Edge spectrum opens
a gap while Wannier bands are still gapless. Other parameters are
fixed at $\gamma_{x}=0.8,\gamma_{y}=0.5$ for all. \label{fig:Extended2_Wannier}}
\end{figure}

The Fourier transformation of $H_{2}$ from real space to momentum space provides us with the corresponding
Bloch Hamiltonian:

\begin{equation}
H_{2}({\bf k})=H_{q}({\bf k})+J_{2}(1-\cos k_{x})\text{\ensuremath{\Gamma_{24}}}+J_{2}\sin k_{x}\Gamma_{23},\label{eq:H2_k}
\end{equation}
where the Gamma matrices are defined as $\Gamma_{ab}=-i\Gamma_{a}\Gamma_{b}$.
Here $\Gamma_{24}$ ($\Gamma_{23}$) is odd (even) under time-reversal
symmetry, and thus we have:

\begin{equation}
\mathcal{T}H_{2}({\bf k})\mathcal{T}^{-1}\neq H_{2}(-{\bf k}),\ \mathcal{T}=K,
\end{equation}
where $\mathcal{T}$ is the time-reversal operator and $K$ represents
complex conjugation. Chiral symmetry $\text{\ensuremath{\mathcal{C}} }$
is also not preserved, while particle-hole symmetry is respected with its operator $\mathcal{P}=\tau_{3}\sigma_{0}K$ given as the production of  $\text{\ensuremath{\mathcal{C}} }$ and $\text{\ensuremath{\mathcal{T}}}$. Here the bulk bands
lose their double degeneracy (Fig. \ref{fig:Extended2}(b)). Crucially,
the additional terms in Eq.~\eqref{eq:H2_k} maintain mirror symmetry
$\mathcal{M}_{x,y}$.

\begin{figure}
\includegraphics[width=1\linewidth]{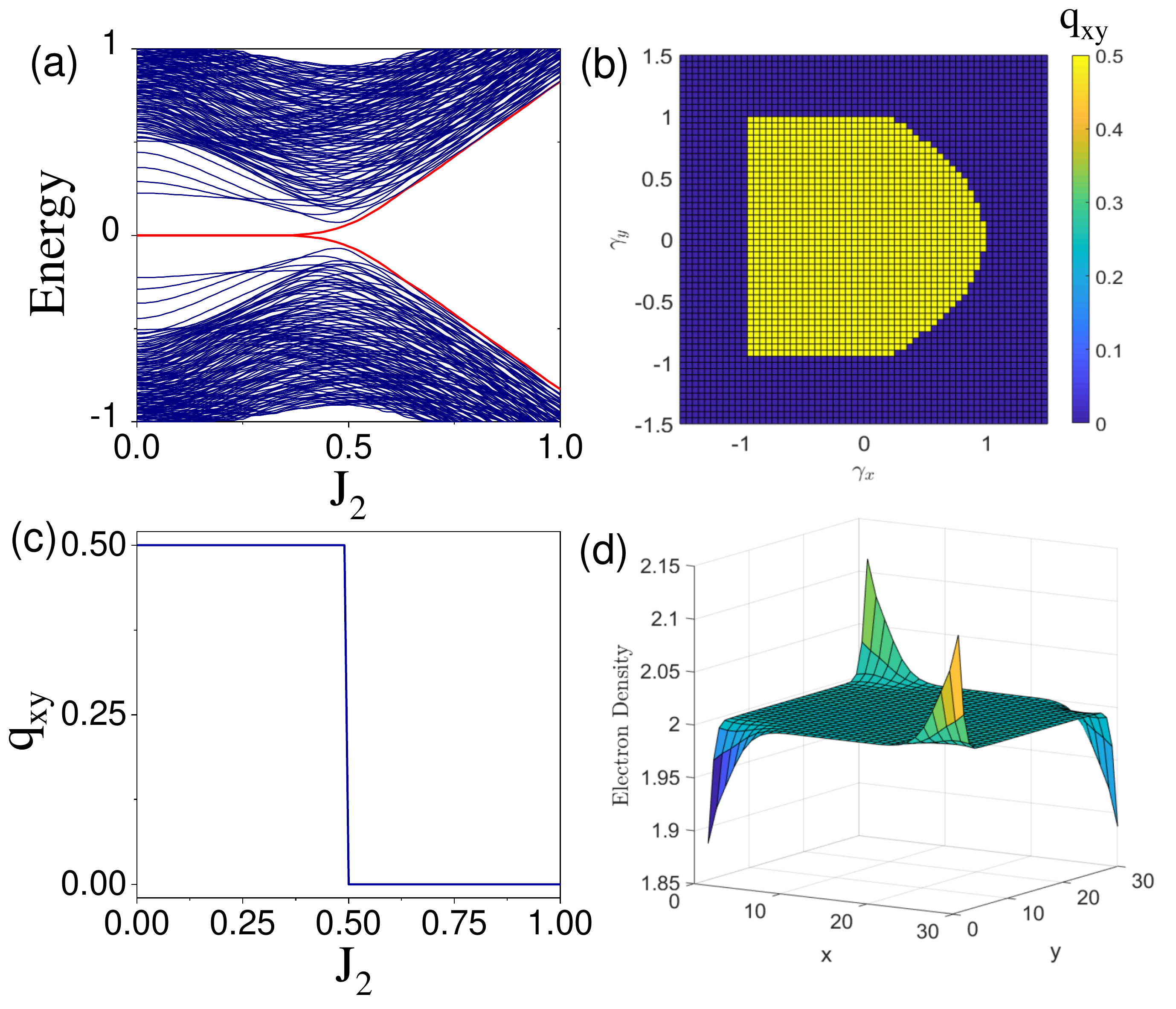}

\caption{The topological phase transition induced by $iJ_{2}$ hopping terms.
(a) Energy spectrum as function of $J_{2}$. We find that the zero-energy
mode splits around $J_{2}=0.5$ due to topological phase transition.
(b) Calculated quadrupole moments $q_{xy}$ for fixed parameter $J_{2}=0.5$.
The nontrivial regions shrink due to topological phase transitions.
(c) Numerically calculated $q_{xy}$ corresponding to (a). Here $q_{xy}$
jumps from one-half to zero around $J_{2}=0.5$. (d) Electron charge
density in the nontrivial phase for $J_{2}=0.2$. Other parameters
are fixed at $\gamma_{x}=0.8,\gamma_{y}=0.5$ for all. Here we set
system at size $L_{x}\times L_{y}=40\times40$ with periodic boundary
for the calculation of $q_{xy}$. \label{fig:Extended2_phase}}
\end{figure}

Two key points should be noted regarding the extended model $H_{2}({\bf k})$.
First, topological equivalence between the edge spectrum and the Wannier bands
may be lost in this model. It has been suggested that the Wannier bands
(e.g. $\nu_{y}(k_{x})$) can be continuously mapped to the
edge spectrum localized at, for example, the $y$-normal boundaries \cite{Neupert18springer}. In other words, the edge spectrum at the $y$-normal boundaries should be topologically equivalent to Wannier bands
$\nu_{y}(k_{x})$.
However, in our case, this is not necessarily true. In Fig. \ref{fig:Extended2_Wannier}(a),
the edge spectrum (red lines in the top panel of Fig. \ref{fig:Extended2_Wannier})
along the $x$ direction closes its gap as we tune $J_{2}$ to approximately $0.5$.
At this point, the Wannier bands close their gap correspondingly.
Slightly increasing $J_{2}$ results in the immediate opening of a gap by the edge spectrum,
while the Wannier bands remain gapless (for a small region
of approximately $0.5\lesssim J_{2}\lesssim0.65$). Although the underlying reasons behind the lack of gaps for the Wannier bands at such a small parameter
region remains unclear, this observation clearly indicates that the topological equivalence
between the Wannier bands and edge spectrum is lost. For $J_{2}$ greater
than $0.65$, the gap of the Wannier bands also opens. A discrepancy
between these two bands can occur when we consider long-range hopping
between unit cells \cite{Yang19arxiv,Yang2019arxiv}. If
we consider the edge spectrum along the $k_{y}$ direction, a similar discrepancy
can be observed. The nested Wilson approach thus loses its validity in describing
the topological phase. This topological
equivalence loss between the Wannier bands and the
edge spectrums consequently indicates that choosing a Wannier
sector to calculate the polarization is illegal. Indeed, the numerically calculated
${\bf p}^{\nu}$ is arbitrary within $[0,1]$ for gapless Wannier bands.
For $J_{2}$ greater than $0.65$, the Wannier bands gap opens and
the Wannier sector polarization gives an incorrect index.

\begin{figure}
\includegraphics[width=1\linewidth]{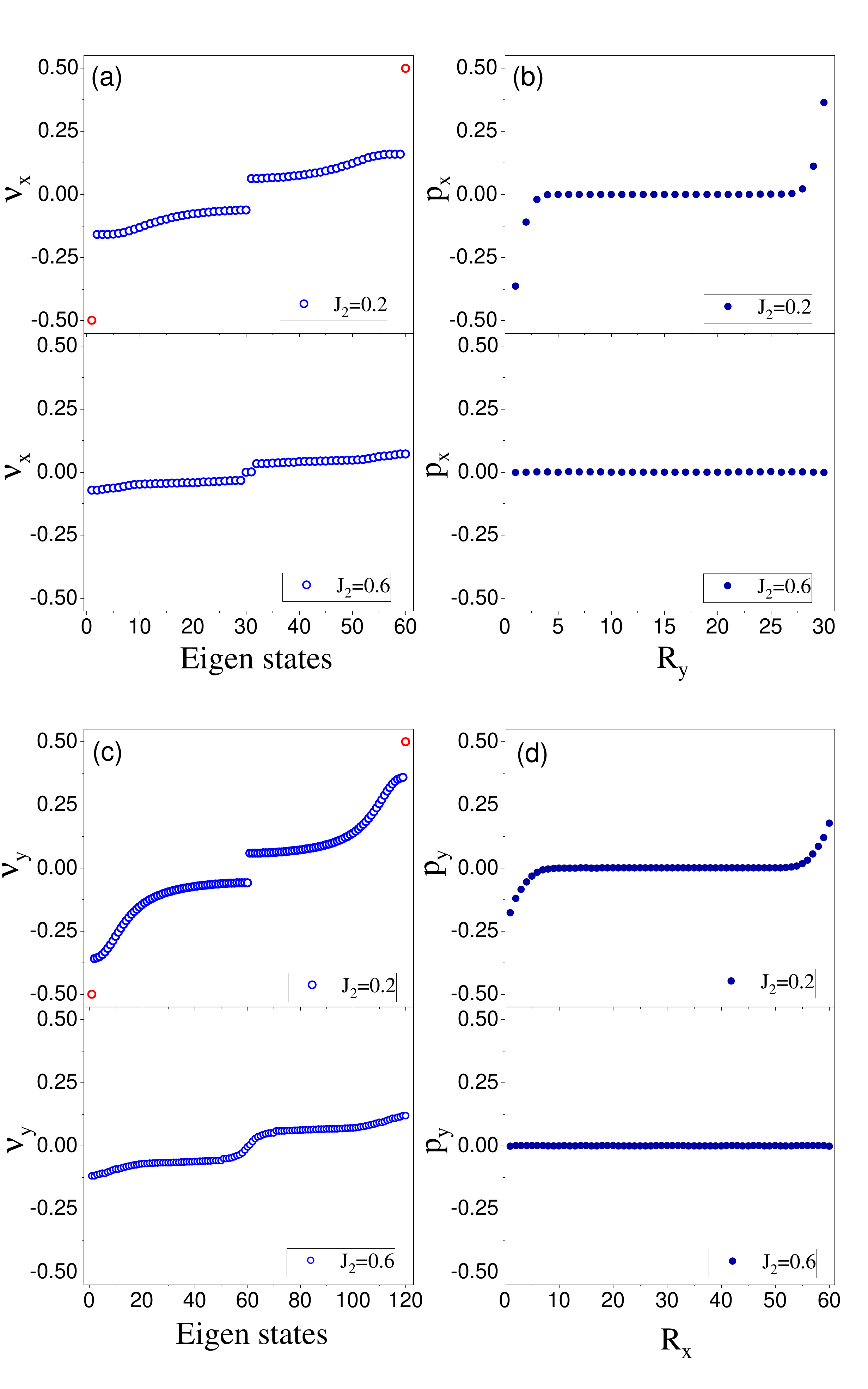}

\caption{Wannier values and edge polarization before ($J_{2}=0.2$) and after
($J_{2}=0.6$) the topological phase transition. (a) $\nu_{x}$ for
different eigen states, (b) edge polarization $p_{x}$ along $R_{y}$,
(c) $\nu_{y}$ for different eigen states, (d) edge polarization $p_{y}$
along $R_{x}$. Other parameters are $\gamma_{x}=0.8,\gamma_{y}=0.5.$
Here the red circles in the upper panel of (a) and (c) are used to
stress the two isolated topological modes with quantized Wannier center
one half. \label{fig:Extended2_polarization}}
\end{figure}

Second, the edge gap close-reopen process is a topological phase transition.
Note that the gap closes at two different $k_{x}$ points. Since the
real phase transition occurs at the sample's boundaries, we focus
on the edge spectrum and ignore the Wannier bands at this point.
In order to explicitly show the phase transition, we employ several signatures
including quadrupole moment $q_{xy}$, fractional corner charges,
and quantized edge polarizations. We compare the variation of these
signatures before and after the edge spectrum closing point to ensure
the occurrence of the topological phase transition. The calculated $q_{xy}$ jumps
from $q_{xy}=1/2$ to $q_{xy}=0$ (Fig. \ref{fig:Extended2_phase}(c)), indicating
that the edge spectrum gap close-reopen process drives the quadrupole
insulator from a topological non-trivial phase to a trivial phase. This
quadrupole moment jump is associated with the disappearance of both
the zero-energy corner modes (Fig. \ref{fig:Extended2_phase}(a))
and the fractional corner charges (see Fig. \ref{fig:Extended2_phase}(d)).
Putting in more detailed words, prior to the phase transition point, the system is gapped with zero-energy
corner states, as shown in Fig. \ref{fig:Extended2_phase}(a). Following
the phase transition point, zero-energy corner modes do not appear
in the trivial gap, which is consistent with the result of $q_{xy}$.
The real-space numerical calculations of $q_{xy}$ prove to be
more effective for the characterization of the quadrupole moments compared to the nested
Wilson loop. It is clear that this topological phase transition will
modify the original phase diagram. In particular, the non-trivial region will
shrink, as shown in Fig. \ref{fig:Extended2_phase}(b).

The edge polarization $p_{x}^{\mathrm{edge,}\pm y}$ and $p_{y}^{\mathrm{edge,}\pm x}$,
together with the states having Wannier values $\pm1/2$, can also aid in the detection of
the topological phase transition. Figure \ref{fig:Extended2_polarization}
depicts the edge polarizations and Wannier centers of the system
at the two sides of the phase transition point. It is evident that
for the non-trivial case where non-zero edge polarizations are present,
both $p_{x}(R_{y})$ and $p_{y}(R_{x})$ are accompanied by two isolated
topological modes with pinned Wannier center $v_{x,y}=\pm1/2$ (red circles). Note that the edge polarizations slightly penetrate into the bulk bands,
yet their integration over half of the lattice width
still results in the quantized value of $\pm e/2$. In comparison, for the trivial
case where the edge polarization is maintained at zero, the states with the quantized
Wannier values disappear.

\section{Robust topological quadrupole phases under mirror symmetry breaking}

\begin{figure}
\includegraphics[width=1\linewidth]{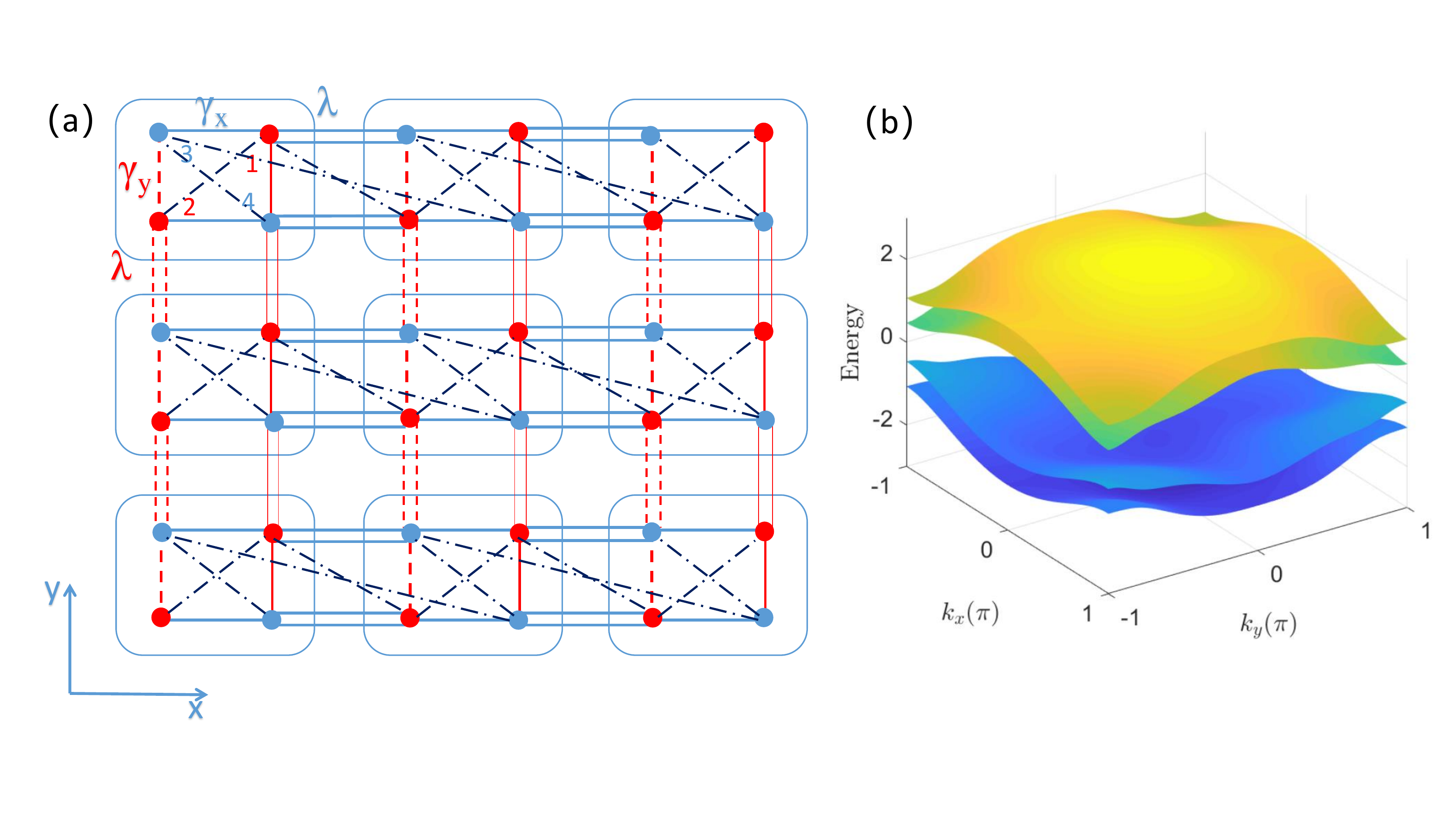}

\caption{(a) Lattice structure of extended BBH model $H_{3}$. The dashed dark
blue lines represent hopping within and between unit cells. Here the
artificial asymmetric hopping breaks mirror symmetry. (b) Energy bands
corresponding to (a) with parameters fixed at $\gamma_{x}=\gamma_{y}=0.5,$
and $J_{3}=0.3$. The band degeneracy is totally lifted. \label{fig:Extended3}}
\end{figure}

Mirror symmetries play an important role for the quantization of
quadrupole moments in the BBH model. It is thus useful to investigate the consequences
of mirror symmetry breaking. To this end, we consider a model that
breaks mirror symmetries explicitly, as shown in Fig. \ref{fig:Extended3}(a).
The generalized model is described as follows:

\begin{alignat}{1}
H_{3} & =H_{0}+\sum_{\mathbf{R}}J_{3}(iC_{\mathbf{R},2}^{\dagger}C_{\mathbf{R},1}-iC_{\mathbf{R},4}^{\dagger}C_{\mathbf{R},3}+\mathrm{H.c.})\nonumber \\
 & +\sum_{\mathbf{R}}J_{3}(iC_{\mathbf{R},1}^{\dagger}C_{\mathbf{R}+\hat{x},2}+iC_{\mathbf{R},3}^{\dagger}C_{\mathbf{R}+\hat{x},4}+\mathrm{H.c.}),\label{eq:H3}
\end{alignat}
where $iJ_{3}$ is the imaginary hopping amplitude. For simplicity, we assume
the same hopping amplitudes within and between unit cells.
Transforming $H_{3}$ into momentum space gives Bloch Hamiltonian:

\begin{equation}
H_{3}({\bf k})=H_{q}({\bf k})+J_{3}\Gamma_{24}-J_{3}(\sin k_{x}\Gamma_{23}-\cos k_{x}\Gamma_{13}).\label{eq:Hk_3}
\end{equation}

The most crucial feature of $H_{3}({\bf k})$ is the breaking of the mirror symmetry,
which is evident from our artificial hopping design depicted in Fig. \ref{fig:Extended3}(a).
Explicitly, we have:

\begin{equation}
\mathcal{M}_{s}H_{3}({\bf k})\mathcal{M}_{s}^{-1}\neq H_{3}(\mathcal{M}_{s}{\bf k}),\ s=x,y.
\end{equation}

In this case, the mirror symmetry operators are $\mathcal{M}_{x}=\tau_{1}\sigma_{3},\ \mathcal{M}_{y}=\tau_{1}\sigma_{1}.$
The rotation symmetry $C_{2}=\mathcal{M}_{x}\mathcal{M}_{y}$ of
$H_{3}({\bf k})$ is preserved, which allows for well-defined quadrupole moments in the vanishing total
bulk polarization \cite{BBH17prb}.
Note that chiral and time-reversal symmetries $\mathcal{C}$ and $\mathcal{T}$
are both absent, thus the discrepancy between the edge spectrum and the Wannier
bands also occurs here. Due to the only preserved particle-hole symmetry,
the system is categorized into the $D$ class, with a classification of $\mathbb{Z}$
in 2D. It is evident from the numerical result in Fig. \ref{fig:Extended3}(b)
that the band degeneracy is totally lifted.

\begin{figure}
\includegraphics[width=1\linewidth]{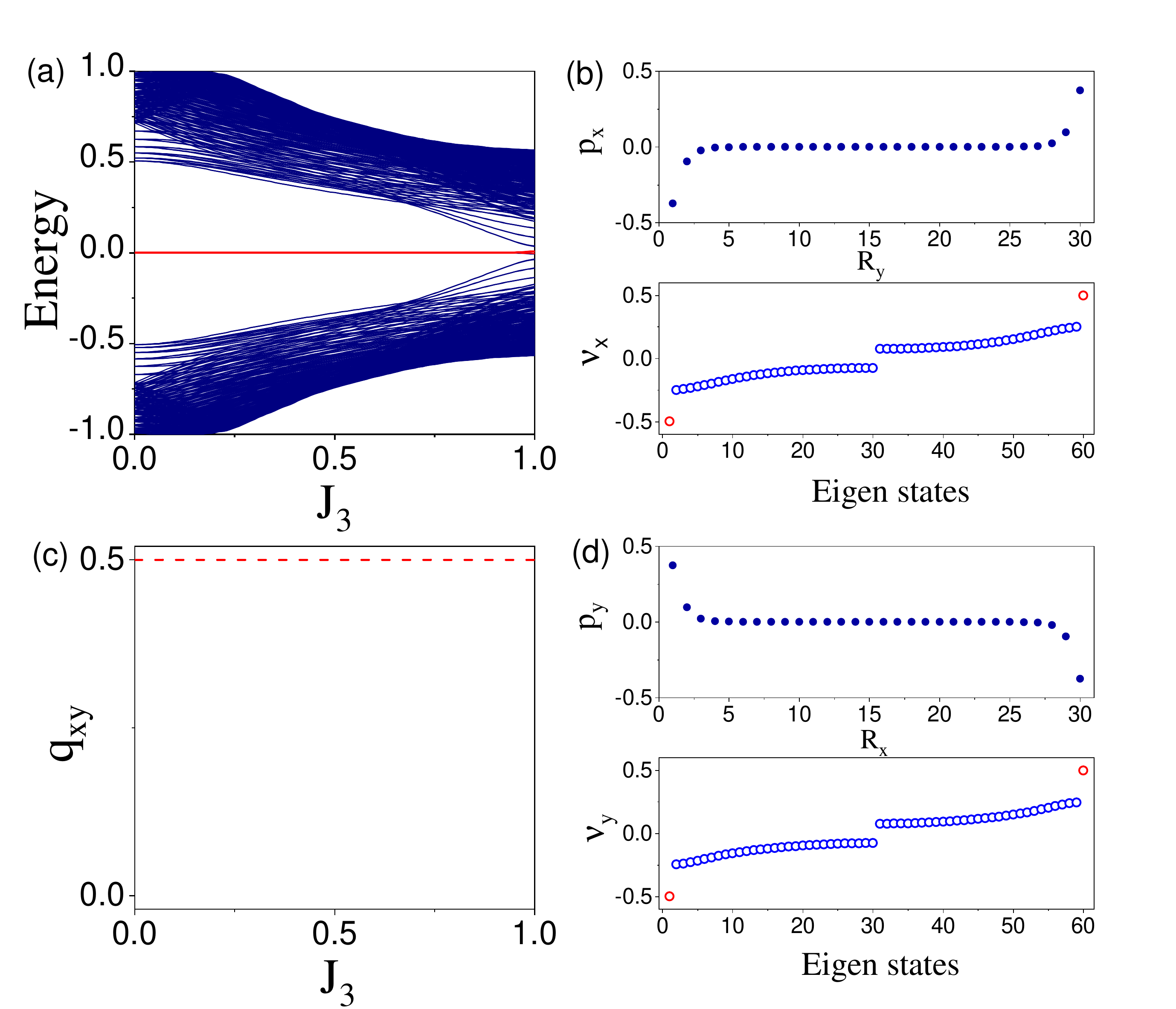}

\caption{(a) Energy spectrum as function of parameter $J_{3}$. (b) Edge polarization
$p_{x}$ along $R_{y}$ (the upper panel) and Wannier center $\nu_{x}$
for different eigen states (the lower panel). (c) Numerically calculated
$q_{xy}$ corresponding to (a) panel. (d) Edge polarization $p_{y}$
along $R_{x}$ (the upper panel) and Wannier center $\nu_{y}$ for different
eigen states (the lower panel). Other parameters are fixed at $\gamma_{x}=\gamma_{y}=0.5,$
and $J_{3}=0.2$. Here we set system size $L_{x}\times L_{y}=40\times40$
with periodic boundary for the calculations of $q_{xy}$. Here the
red circles in the lower panel of (b) and (d) are used to stress the
two isolated topological modes with quantized Wannier center one half.
\label{fig:Extended3-1}}
\end{figure}

A crucial result of the generalized model $H_{3}({\bf k})$
is the persistence of the topological quadrupole phases despite the breakdown of the mirror
symmetries. In the original BBH model \cite{Benalcazar17Science,BBH17prb},
the quantization of the quadrupole moments is protected by the combination
of the mirror symmetries ($H_{3}({\bf k})$ does not respect $C_{4}$
symmetry). In the $H_{3}({\bf k})$ model, the robustness of topological quadrupole phases can be demonstrated in several ways.
First, it is always valid to calculate quadrupole
moments $q_{xy}$. In Fig. \ref{fig:Extended3-1}(c), the numerical results
indeed show that for the existing parameters, $q_{xy}$ quantizes at $1/2$ for increasing $J_{3}$. Due to the lack of mirror symmetries,
the Wannier sector polarization ${\bf p}^{\nu}$, however, is not
quantized. Second, zero-energy corner modes are present. The red line in the spectrum
of Fig. \ref{fig:Extended3-1}(a) indicates four-fold
degenerate zero-energy modes whose wave functions are sharply localized
at the corners of the sample. Moreover, the corner
charges are quantized at $\pm e/2$. Third, we found that non-trivial
quadrupole phases also correspond to quantized edge polarizations.
The lower panels of Figs. \ref{fig:Extended3-1}(b) and \ref{fig:Extended3-1}(d)
show that the two topological states located on the edge exhibit half-integer Wannier
values, while other states are distributed over the bulk. The edge polarization
$p_{x}(R_{y})$ (or $p_{y}(R_{x})$) has a non-zero value close to the
edge of the sample. Despite not being highly distributed at the edge
sites, the integrated polarization over half of the lattice width
still results in the quantized edge polarization of $p_{x}^{\mathrm{edge},\pm y}=\pm e/2$
($p_{y}^{\mathrm{edge},\pm x}=\pm e/2$). The above observations
unambiguously determine the presence of the topological quadrupole phases.

The robustness of $q_{xy}$ under mirror-symmetry
breaking can be explained as follows. Although mirror symmetry is used to construct
the topological quadrupole insulators, its protection is not necessary
for their existence. Once the mirror symmetry is broken, the
robust states located at the corners still remain \cite{Langbehn17prl,Trifunovic19prx},
and more generally, the existence of corner modes even does not require
crystalline symmetry \cite{zhang19arxiv,You2019arXiv}. The corner
states are associated with the quantized corner charges $\pm e/2$, and
cannot be changed adiabatically. This may consequently account for the quantization
of $q_{xy}$.

\section{Quantized quadrupole moments against disorders}

In this section, we investigate the disorder effect on the BBH model. Essentially, the
topological quadrupole insulator is a topological crystalline
insulator, and the presence of disorder can subsequently test the robustness of
the topological phase. Furthermore, the involvement of the disorder may induce
more interesting topological phases, such as the topological Anderson
insulators \cite{LiJian09prl}.

\begin{figure}
\includegraphics[width=1\linewidth]{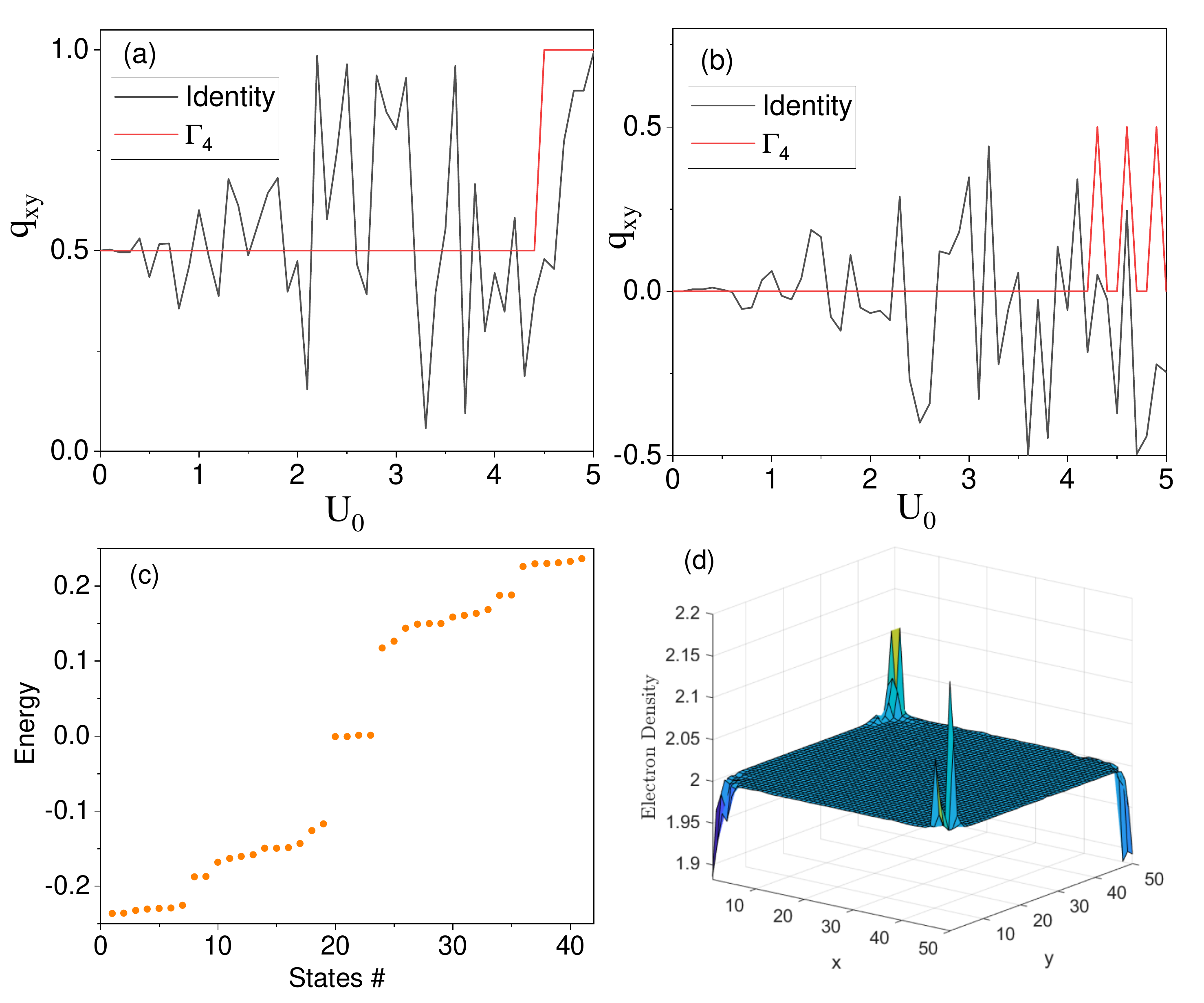}

\caption{Quadrupole moments in the presence of disorder. Upper panels (a) and
(b) plot quadrupole moments $q_{xy}$ as functions of disorder strength
$U_{0}$. In topological nontrivial case $\gamma_{x}=0.8$, $\gamma_{y}=0.5$
and trivial case with $\gamma_{x}=1.5$, $\gamma_{y}=0.5$. In panels
(a,b) we use periodic boundary condition. (c) Eigen states and (d)
electron charge density in the presence of $\Gamma_{4}$ type disorder
with strength $U_{0}=3.0$ for the nontrivial case $\gamma_{x}=0.8$,
$\gamma_{y}=0.5$. In panels (c,d) we use open boundary condition.
The system size is $L_{x}\times L_{y}=50\times50$, and we only take
one disorder configuration for all. \label{fig:Extended4}}
\end{figure}

We first consider a simple on-site type of disorder of the form
$V_{\mathrm{dis}}=V({\bf R})I_{4\times4}$ at each lattice site ${\bf R}$.
Here $I_{4\times4}$ is an identity matrix and the random onsite potential,
$V({\bf R})$, distributes uniformly in the interval $[-U_{0}/2,U_{0}/2]$
for disorder strength $U_{0}$. Figure \ref{fig:Extended4} shows that
for relatively small $U_{0}$ values, the quadrupole moments $q_{xy}$ are well-quantized.
This is a reasonable observation as the disorder is
weak and can be considered as a series of small perturbations. In addition, for large $U_{0}$ (compared to the bulk gap), $q_{xy}$ exhibits a highly fluctuating trend.
If we take the averages across the disorder, the situation improves.
More specifically, the quantization of the quadrupole moments recovers under the condition
of ``averaged mirror symmetry'' \cite{Benalcazar17Science}. However, for much larger
values of $U_{0}$, maintaining the quantization of $q_{xy}$ proves impossible.

An interesting phenomenon is observed when we consider another type of disorder,
namely, the hopping disorder of the form $V({\bf R})\Gamma_{4}$.
Fig. \ref{fig:Extended4} presents the specific parameter configuration
used for this case. Results show that the quadrupole moments are
well-quantized, regardless of the large disorder strength and lack of disorder average.
The quantization of the quadrupole moments is demonstrated by the zero-energy corner modes in the bulk gap (Fig. \ref{fig:Extended4}(c)) and the quantized corner charges $\pm e/2$
(Fig. \ref{fig:Extended4}(d)). This robustness may be explained
as follows. From a 1D point of view, the BBH model can
be decomposed into two uncoupled SSH models along the $x$ and $y$ directions.
Expressed in terms of $\Gamma$ matrices, the $V({\bf R})\Gamma_{4}$-type disorder respects the chiral symmetry of the SSH model along the $x$
direction. Thus, the topology of each SSH model is still well-defined,
regardless of the presence of such a strong disorder \cite{Shem14prl}. As the
disorder strength approaches infinity, we find that the disorder effect
dominates and the system becomes trivial. The phase transition from
a non-trivial to trivial phase occurs (Fig. \ref{fig:Extended4}(a))
and the phase diagram is modified. Interestingly, the high-order
topological Anderson-type phase can also appear due to the $\Gamma_{4}$-type disorder \cite{LiCA19}. As the central theme of this paper is the quantization of quadrupole moments, here we just focus
on the robustness of $q_{xy}$, which remains even under the strong $\Gamma_{4}$ type of disorder
where crystalline symmetry is broken down.

\section{discussion and conclusions}

On the one hand, non-spatial symmetries are not necessary to protect
quantization of quadrupole moments, while several
generalized models demonstrating that non-spatial symmetry breaking can lead to
richer topological phases in the system. On the other hand, although
the BBH model is constructed based on a combination of mirror symmetries,
when mirror symmetry is broken the quantized quadrupole moments and topologically protected states located
at the corners still remain.
Thus, the underlying roles played
by different symmetries within higher multipole insulators should be further investigated in order to determine their specific impacts on the system.

In summary, we investigate the topological quadrupole phases and consequences
of symmetry breaking in the generalized BBH models. We summarize
the properties of several generalized models (including
the original BBH model) in Table \ref{table:parameters}. Results from the symmetry analysis indicate that only $C_{2}$ symmetry (inversion symmetry) is required to ensure well-defined
quadrupole moments, and quantized quadrupole moments $q_{xy}$ can persist
even when mirror symmetry is broken. Besides, we note that nested Wilson
loop approach is not suitable to characterize the high-order topology
of the generalized models and the topological equivalence between
Wannier bands and edge spectrum could be lost; Interestingly, both the quantized
quadrupole moments and the topologically protected boundary signatures
continue to exist despite the presence of a unique type of disorder.
Our results may give a clue to search for more \textquoteleft robust\textquoteright{}
topological invariant of high-order topological phases.
\begin{widetext}
\begin{table}[H]
\centering

\caption{Symmetries and properties of the quantized quadrupole insulator models. }

\label{table:parameters}

\begin{tabular}{cccccccccc}
\toprule
 & \quad{}$\mathcal{M}_{x}$\quad{}  & \quad{}$\mathcal{M}_{y}$\quad{}  & \quad{}$C_{2}$\quad{}  & \quad{}$C_{4}$\quad{}  & \quad{}$\mathcal{C}$\quad{}  & \quad{}$\mathcal{\mathcal{T}}$\quad{}  & \quad{}$\mathcal{P}$\quad{}  & $q_{xy}=0,1/2$  & Quantized $p^{\mathrm{edge}}$\tabularnewline
\midrule
$H_{0}({\bf k})$  & \textbf{\LARGE{}{}$\checkmark$}{\LARGE{} } & \textbf{\LARGE{}{}$\checkmark$}{\LARGE{} } & \textbf{\LARGE{}{}$\checkmark$}{\LARGE{} } & \textbf{\LARGE{}{}$\checkmark$}{\LARGE{} } & \textbf{\LARGE{}{}$\checkmark$}{\LARGE{} } & \textbf{\LARGE{}{}$\checkmark$}{\LARGE{} } & \textbf{\LARGE{}{}$\checkmark$}{\LARGE{} } & \textbf{\LARGE{}{}$\checkmark$}{\LARGE{} } & \textbf{\LARGE{}{}$\checkmark$}\tabularnewline
\midrule
$H_{1}({\bf k})$  & \textbf{\LARGE{}{}$\checkmark$}{\LARGE{} } & \textbf{\LARGE{}{}$\checkmark$}{\LARGE{} } & \textbf{\LARGE{}{}$\checkmark$}{\LARGE{} } & \textbf{\LARGE{}{}$\checkmark$}{\LARGE{} } & \textbf{\LARGE{}{}$\times$}{\LARGE{} } & \textbf{\LARGE{}{}$\checkmark$}{\LARGE{} } & \textbf{\LARGE{}{}$\times$}{\LARGE{} } & \textbf{\LARGE{}{}$\checkmark$}{\LARGE{} } & \textbf{\LARGE{}{}$\checkmark$}\tabularnewline
\midrule
$H_{2}({\bf k})$  & \textbf{\LARGE{}{}$\checkmark$}{\LARGE{} } & \textbf{\LARGE{}{}$\checkmark$}{\LARGE{} } & \textbf{\LARGE{}{}$\checkmark$}{\LARGE{} } & \textbf{\LARGE{}{}$\times$}{\LARGE{} } & \textbf{\LARGE{}{}$\times$}{\LARGE{} } & \textbf{\LARGE{}{}$\times$}{\LARGE{} } & \textbf{\LARGE{}{}$\checkmark$}{\LARGE{} } & \textbf{\LARGE{}{}$\checkmark$}{\LARGE{} } & \textbf{\LARGE{}{}$\checkmark$}\tabularnewline
\midrule
$H_{3}({\bf k})$  & \textbf{\LARGE{}{}$\times$}{\LARGE{} } & \textbf{\LARGE{}{}$\times$}{\LARGE{} } & \textbf{\LARGE{}{}$\checkmark$}{\LARGE{} } & \textbf{\LARGE{}{}$\times$}{\LARGE{} } & \textbf{\LARGE{}{}$\times$}{\LARGE{} } & \textbf{\LARGE{}{}$\times$}{\LARGE{} } & \textbf{\LARGE{}{}$\checkmark$}{\LARGE{} } & \textbf{\LARGE{}{}$\checkmark$}{\LARGE{} } & \textbf{\LARGE{}{}$\checkmark$}\tabularnewline
\bottomrule
\end{tabular}
\end{table}
\end{widetext}

\section{Acknowledgments}

C.-A. Li thanks B. Fu, Z.-A. Hu, J. Li, and S.-Q. Shen for helpful
discussions, and acknowledges H.-Z Lu at SUSTech. China for hospitality
at the initial stage of this project. This work is supported by foundation
of Westlake University and the Natural Science Foundation of Zhejiang Province under Grant No. Q20A04005.

\appendix

\section{The effective model}

In this appendix, we derive an effective Hamiltonian from BBH model.
For the chosen parameters, the bulk gap of BBH model closes at $(\pi,\pi)$
point when $\gamma_{x}=\gamma_{y}=1$. Expanding Eq.~\eqref{eq:H_q}
around this gap closing point to the second order, we have an effective
model

\begin{alignat}{1}
H_{\mathrm{eff}}({\bf k}) & =\left(\begin{array}{cc}
0 & Q\\
Q^{\dagger} & 0
\end{array}\right),\nonumber \\
Q & \equiv-m_{x}\sigma_{0}-ik_{x}\sigma_{3}-im_{y}\sigma_{2}-ik_{y}\sigma_{1},
\end{alignat}
where $m_{x}\equiv1-\gamma_{x}-k_{x}^{2}/2$, and $m_{y}\equiv1-\gamma_{y}-k_{y}^{2}/2$.
Note that this effective Hamiltonian inherits all symmetries of the
original BBH model. After proper rotation of gamma matrices, we arrive
at

\begin{equation}
H_{\mathrm{eff}}({\bf k})=m_{x}\beta+\alpha\cdot{\bf p},\label{eq:Dirac Eq.}
\end{equation}
where the Dirac matrices are defined as $\beta=\tau_{3}\sigma_{0},$
and $\alpha_{i}=\tau_{1}\sigma_{i}.$ The ``momentums'' in Eq.~\eqref{eq:Dirac Eq.}
are defined as
\begin{equation}
{\bf p}\equiv(k_{y},m_{y},k_{x}).
\end{equation}
Here we recover the fact that this effective model Eq.~\eqref{eq:Dirac Eq.}
is exactly Dirac equation in two dimensions \cite{SQS}, whereas one
of the momentums is replaced by mass term $m_{y}$. Tracing back the
relation between $\Gamma_{i}$ matrices before rotation and $\{\beta,{\bf \alpha}\}$
Dirac matrices in Eq.~\eqref{eq:Dirac Eq.}, we find that $\Gamma_{4}\rightarrow\beta$.
This means that the disorder of the form $V({\bf R})\Gamma_{4}$ modifies
the mass term of Dirac Hamiltonian ~\eqref{eq:Dirac Eq.}. Besides,
it explains why the quadrupole is still quantized when strong disorder
is applied, as shown in Fig. \ref{fig:Extended4}, to some extent.

Actually, Eq.~\eqref{eq:Dirac Eq.} describes the same physics as
quantum spin Hall insulator with an anisotropic edge gap term. This point would
be transparent if the Dirac matrices are rotated by a unitary transformation
\begin{equation}
\left(\begin{array}{c}
\alpha'_{2}\\
\beta'
\end{array}\right)=\frac{1}{\sqrt{2}}\left(\begin{array}{cc}
1 & -1\\
1 & 1
\end{array}\right)\left(\begin{array}{c}
\alpha_{2}\\
\beta
\end{array}\right).\label{eq:Unitary Trans.}
\end{equation}
For simplicity, let us set $\gamma_{x}=\gamma_{y}=\gamma$. With the
help of Eq.~\eqref{eq:Unitary Trans.}, the effective model is transformed
to

\begin{alignat}{1}
H_{\mathrm{eff}}({\bf k}) & =m({\bf k})\beta'+k_{x}\alpha_{3}+k_{y}\alpha_{1}+\Delta({\bf k})\alpha'_{2},\label{eq:rotated}\\
m({\bf k}) & \equiv\left[\sqrt{2}(1-\gamma)-\frac{k^{2}}{2\sqrt{2}}\right],\Delta({\bf k})\equiv\frac{k_{x}^{2}-k_{y}^{2}}{2\sqrt{2}}.
\end{alignat}
From the mass term $m({\bf k})$, the system is topologically nontrivial
when $\gamma<1$, which is consistent with BBH model. Note that the
Dirac matrices after rotation still obey Clifford algebra. If $\Delta({\bf k})=0$,
which is only satisfied if $k_{x}^{2}=k_{y}^{2}$, the effective Hamiltonian
describes quantum spin Hall insulator. Generally, $\Delta({\bf k})$
acts as a mass term to gap helical edge modes as it anticommutes with
other Dirac matrices in the Eq.~\eqref{eq:rotated}, and it creates
a domain wall profile since it changes sign at adjacent two edges.
In this sense, the BBH model is equivalent to a quantum spin Hall
insulator with an anisotropic edge gap term.

\section{Calculation details}

In this appendix, we show the details about the calculation of nested
Wilson loop approach, edge polarization, and etc. \cite{Benalcazar17Science,BBH17prb}.

\subsection{Wannier bands}

The nested Wilson loop approach is based on Wilson loop, thus let
us start with the construction of Wilson loop. The Wilson loop operator
parallel to the $y$ direction is constructed as

\begin{equation}
\hat{P}_{y,{\bf k}}=P_{N_{y}\delta k_{y}+k_{y}}P_{(N_{y}-1)\delta k_{y}+k_{y}}\cdots P_{\delta k_{y}+k_{y}}P_{k_{y}},\label{eq:Wilsonloop}
\end{equation}
where each projection operator is defined as $P_{m\delta k_{y}+k_{y}}\equiv\sum_{n\in N_{\mathrm{occ}}}|u_{k_{x},m\delta k_{y}+k_{y}}^{n}\rangle\langle u_{k_{x},m\delta k_{y}+k_{y}}^{n}|$
with $|u_{k_{x},m\delta k_{y}+k_{y}}^{n}\rangle$ being the $n$-th
eigen state of occupied bands at point $(k_{x},m\delta k_{y}+k_{y})$,
and $m$ is an integer taking values from $\{1,2,\cdots,N_{y}\}$.
The projection method can avoid the arbitrary phase problem in numerical
realizations. Here $N_{y}$ is the number of unit cells, $n$ is the
band index, and $N_{\mathrm{occ}}$ is the number of occupied bands.
Note that $\hat{P}_{y,{\bf k}}$ has dimension of $N$ now with $N$
being the total bands number. After projection onto the occupied bands
at base point ${\bf k}$, there is $N_{\mathrm{occ}}\times N_{\mathrm{occ}}$
matrix $\mathcal{W}_{y,{\bf k}}$ that defines a Wannier Hamiltonian
$H_{\mathcal{W}_{y}}({\bf k})$ from the relation $\mathcal{W}_{y,{\bf k}}=\exp[iH_{\mathcal{W}_{y}}({\bf k})]$.
The eigen values of $H_{\mathcal{W}_{y}}({\bf k})$ give the Wannier
bands $2\pi\nu_{y}(k_{x})$ associated with eigen states $|\nu_{y,{\bf k}}^{j}\rangle$,
$j\in\{1,2,\cdots,N_{\mathrm{occ}}\}$. Similar procedure is applicable
for the construction of $\mathcal{W}_{x,{\bf k}}$.

\subsection{Wannier sector polarization}

These Wannier bands can carry their own topology and have associated
Berry phases. If the Wannier bands are gapped, then one can chose
a Wannier sector and construct nested Wilson loop $\mathcal{\widetilde{W}}_{x,k_{y}}$
by the similar projection procedures as before, which gives rise to
Wannier sector polarization $p_{x}^{v_{y}}$. To this end, we need
to define the Wannier basis

\begin{equation}
|w_{y,{\bf k}}^{\ell}\rangle=\sum_{n}^{N_{\mathrm{occ}}}|u_{{\bf k}}^{n}\rangle[\nu_{y,{\bf k}}^{\ell}]^{n},
\end{equation}
where $\ell$ takes value from the set $\{1,2,\cdots,N_{\mathrm{ws}}\}$,
and $N_{\mathrm{ws}}$ is the dimension of chosen Wannier sector.
Here $[\nu_{y,{\bf k}}^{\ell}]^{n}$ is the $n$-th component of eigen
state $|\nu_{y,{\bf k}}^{\ell}\rangle$. The nested Wilson loop operator
parallel along the $x$ direction is constructed as
\begin{equation}
\hat{\mathcal{P}}_{x,{\bf k}}=\mathcal{P}_{N_{x}\delta k_{x}+k_{x}}\mathcal{P}_{(N_{x}-1)\delta k_{x}+k_{x}}\cdots\mathcal{P}_{\delta k_{x}+k_{x}}\mathcal{P}_{k_{x}},
\end{equation}
where each projection operator is defined as $\mathcal{P}_{q\delta k_{x}+k_{x}}\equiv\sum_{\ell=1}^{N_{\mathrm{ws}}}|w_{y,q\delta k_{x}+k_{x},k_{y}}^{\ell}\rangle\langle w_{y,q\delta k_{x}+k_{x},k_{y}}^{\ell}|$
where $q$ is also an integer taking values from the set $\{1,2,\cdots,N_{x}\}$.
Note that $\hat{\mathcal{P}}_{x,{\bf k}}$ also has dimension of $N$
now. After projection onto the Wannier basis of chosen sector, there
is $N_{\mathrm{ws}}\times N_{\mathrm{ws}}$ matrix $\mathcal{\widetilde{W}}_{x,k_{y}}$
that defines a nested Wilson loop. The polarization of Wannier bands
is given as
\begin{equation}
p_{x}^{\nu_{y}}=\frac{1}{2\pi iN_{y}}\sum_{k_{y}}\log\det\mathcal{\widetilde{W}}_{x,k_{y}}.
\end{equation}
Similar procedure can be carried out for the construction of $p_{y}^{\nu_{x}}$.

\subsection{Edge polarization}

The edge polarization is exhibited on ribbon samples with finite width
but infinite length \cite{BBH17prb}. It is a useful signature to
exhibit high-order topology in quadrupole insulators. Let us assume
the ribbon is infinite along the $x$ direction. First, we treat the width
along $y$ as inner degree of freedom and get Wannier band $v_{k_{x}}^{s}$using
the projection method described as above. The associated wave functions
are $|\nu_{k_{x}}^{s}\rangle$ where $s\in\{1,2,\cdots,N_{\mathrm{tot}}\}$
where $N_{\mathrm{tot}}=N_{y}\times N_{\mathrm{occ}}$. Second, we
construct hybrid Wannier function as
\begin{equation}
|\Psi_{R_{x}}^{s}\rangle=\frac{1}{\sqrt{N_{x}}}\sum_{k_{x}}\sum_{n=1}^{N_{\mathrm{tot}}}[v_{k_{x}}^{s}]^{n}e^{-ik_{x}R_{x}}\varUpsilon_{n,k_{x}}^{\dagger}|0\rangle,
\end{equation}
where $\varUpsilon_{n,k_{x}}$ is the basis that diagonalizes the
Hamiltonian. One can check that $|\Psi_{R_{x}}^{s}\rangle$ is a complete
basis such that $\langle\Psi_{R_{x}}^{s}|\Psi_{R_{x}}^{s'}\rangle=\delta_{ss'}.$
Third, we consider the probability distribution of the Wannier function
along the $y$ direction
\begin{alignat}{1}
\rho^{s}(R_{y}) & =\sum_{R'_{x},\zeta}\langle\Psi_{R_{x}}^{s}|\phi_{R_{x}^{'}}^{R_{y,\zeta}}\rangle\langle\phi_{R_{x}^{'}}^{R_{y,\zeta}}|\Psi_{R_{x}}^{s}\rangle,\nonumber \\
|\phi_{R_{x}^{'}}^{R_{y,\zeta}}\rangle & =\sum_{k_{x}}e^{-ik_{x}R_{x}^{'}}C_{k_{x},R_{y},\zeta}^{\dagger}|0\rangle,
\end{alignat}
where $\zeta$ denotes the inner degree of freedom at each unit cell.
Finally, the edge polarization along the $x$ direction is
\begin{equation}
p_{x}(R_{y})=\sum_{s=1}^{N_{\mathrm{tot}}}\rho^{s}(R_{y})v_{k_{x}}^{s}.
\end{equation}
The total edge polarization $p_{x}^{\mathrm{edge}}$ is defined as
summation of $p_{x}(R_{y})$ over half of the width along the $y$ direction,
for example

\begin{equation}
p_{x}^{\mathrm{edge,}-y}=\sum_{R_{y}=1}^{N_{y}/2}p_{x}(R_{y}).
\end{equation}

\subsection{Corner charges}

The corner charge is also a direct signature of high-order topology
of quadrupole insulators. The local charge density is

\begin{equation}
\rho(R_{x},R_{y})\equiv e\sum_{n=1}^{N_{\mathrm{occ}}}\sum_{\zeta=1}^{4}\left|u^{n}(R_{x},R_{y},\zeta)\right|^{2},
\end{equation}
where $u^{n}(R_{x},R_{y},\zeta)$ is the component of $n$-th eigen
state $|u^{n}\rangle$. The corner charge is defined as the summation
of charge density over a quarter of the sample, for instance,
\begin{equation}
Q^{\mathrm{corner,-x,-y}}=\sum_{R_{x}=1}^{N_{x}/2}\sum_{R_{y}=1}^{N_{y}/2}[\rho(R_{x},R_{y})-2e].
\end{equation}
Note that here we need to eliminate contributions from the atomic
charge $2e$.

\subsection{Quadrupole moments}

Classically, the quadrupole moments are defined as
\begin{equation}
q_{ij}=\int d^{3}r\rho(r)r_{i}r_{j},
\end{equation}
where $\rho(r)$ is the charge density. The quadrupole moments will
manifest observable electromagnetic properties such as corner charges.
BBH proposed quantum mechanical crystalline insulator that holds only
bulk quadrupole moments, and the defining constraint of the quadrupole
insulator is

\begin{equation}
q_{xy}=|p_{x}^{\mathrm{edge}}|=|p_{y}^{\mathrm{edge}}|=|Q^{\mathrm{corner}}|.
\end{equation}
In the presence of certain symmetries, the quadrupole moments will
be quantized, and they are related to Wannier sector polarizations
as
\begin{equation}
q_{xy}\overset{\mathcal{M}_{x},\mathcal{M}_{y}}{=}2p_{x}^{\nu_{y}^{-}}p_{y}^{\nu_{x}^{-}}.
\end{equation}

\subsection{Phases of BBH model}

The BBH model Eq.~\eqref{eq:H_q} is a concrete minimal model that holds
quantized bulk quadrupole moments. Its bulk bands are gapped unless
$\gamma_{s}/\lambda=\pm1$ ($s=x,y$). Hence it is an insulator at
half-filling. The model gives rise to topological quadrupole moments
protected by mirror symmetries in principle. The non-spatial symmetries
preserved are chiral symmetry, time-reversal symmetry, and particle-hole
symmetry, although they are not necessarily needed to quantize quadrupole
moments. The topological phase in electric quadrupole insulators is
characterized by quantized quadrupole moments $q_{xy}=0,1/2$, which
induce corner charge $Q^{\mathrm{corner}}$ and edge polarization
$p^{\mathrm{edge}}$ of its equal magnitude. Explicitly, the quantized
quadrupole moments are formulated via the nested Wilson loops approach.
Sitting on the basis that Wannier bands and boundary spectrum are
topologically equivalent, the topological quadrupole phase is characterized
by Wannier sector polarization ${\bf p}^{\nu}\equiv(p_{y}^{\nu_{x}^{-}},p_{x}^{\nu_{y}^{-}})$.
The nontrivial topological quadrupole phase constrains parameter region
$|\gamma_{s}/\lambda|<1$ for $s=x,y$.

\begin{thebibliography}{0}%
\makeatletter
\providecommand \@ifxundefined [1]{%
 \@ifx{#1\undefined}
}%
\providecommand \@ifnum [1]{%
 \ifnum #1\expandafter \@firstoftwo
 \else \expandafter \@secondoftwo
 \fi
}%
\providecommand \@ifx [1]{%
 \ifx #1\expandafter \@firstoftwo
 \else \expandafter \@secondoftwo
 \fi
}%
\providecommand \natexlab [1]{#1}%
\providecommand \enquote  [1]{``#1''}%
\providecommand \bibnamefont  [1]{#1}%
\providecommand \bibfnamefont [1]{#1}%
\providecommand \citenamefont [1]{#1}%
\providecommand \href@noop [0]{\@secondoftwo}%
\providecommand \href [0]{\begingroup \@sanitize@url \@href}%
\providecommand \@href[1]{\@@startlink{#1}\@@href}%
\providecommand \@@href[1]{\endgroup#1\@@endlink}%
\providecommand \@sanitize@url [0]{\catcode `\\12\catcode `\$12\catcode
  `\&12\catcode `\#12\catcode `\^12\catcode `\_12\catcode `\%12\relax}%
\providecommand \@@startlink[1]{}%
\providecommand \@@endlink[0]{}%
\providecommand \url  [0]{\begingroup\@sanitize@url \@url }%
\providecommand \@url [1]{\endgroup\@href {#1}{\urlprefix }}%
\providecommand \urlprefix  [0]{URL }%
\providecommand \Eprint [0]{\href }%
\providecommand \doibase [0]{http://dx.doi.org/}%
\providecommand \selectlanguage [0]{\@gobble}%
\providecommand \bibinfo  [0]{\@secondoftwo}%
\providecommand \bibfield  [0]{\@secondoftwo}%
\providecommand \translation [1]{[#1]}%
\providecommand \BibitemOpen [0]{}%
\providecommand \bibitemStop [0]{}%
\providecommand \bibitemNoStop [0]{.\EOS\space}%
\providecommand \EOS [0]{\spacefactor3000\relax}%
\providecommand \BibitemShut  [1]{\csname bibitem#1\endcsname}%
\let\auto@bib@innerbib\@empty
\end{thebibliography}%


\begin{thebibliography}{10}
\bibitem{Resta94rmp}R. Resta, \textit{Macroscopic polarization in
crystalline dielectrics: the geometric phase approach}, Rev. Mod.
Phys. \textbf{66}, 899 (1994).

\bibitem{Resta98prl}R. Resta, \textit{Quantum-mechanical position
operator in extended systems}, Phys. Rev. Lett. \textbf{80}, 1800
(1998).

\bibitem{King93prb}R. D. King-Smith and D. Vanderbilt, \textit{Theory
of polarization of crystalline solids}, Phys. Rev. B \textbf{47},
1651 (1993).

\bibitem{XiaoD10rmp}D. Xiao, Ming-Che Chang, and Qian Niu, \textit{Berry
phase effects on electronic properties}, Rev. Mod. Phys. \textbf{82},
1959 (2010).

\bibitem{Kane10rmp}M. Z. Hasan and C. L. Kane, \textit{Colloquium:
Topological insulators,} Rev. Mod. Phys. \textbf{82}, 3045 (2010).

\bibitem{QiXL11rmp}X.-L. Qi and S.-C. Zhang, \textit{Topological
insulators and superconductors}, Rev. Mod. Phys. \textbf{83}, 1057
(2011).

\bibitem{SQS}S.-Q. Shen, \textit{Topological Insulators: Dirac Equation
in Condensed Matter}, 2nd ed. (Springer, Singapore, 2017).

\bibitem{BernevigBook}B. A. Bernevig and T. L. Hughes, \textit{Topological
insulators and topological superconductors} (Princeton University
Press, 2013).

\bibitem{Thouless83prb}D. J. Thouless, \textit{Quantization of particle
transport}, Phys. Rev. B \textbf{27}, 6083 (1983).

\bibitem{FuL06prb}L. Fu and C. L. Kane, \textit{Time reversal polarization
and a $Z_{2}$ adiabatic spin pump}, Phys. Rev. B \textbf{74}, 195312
(2006).

\bibitem{SSH79prl}W. P. Su, J. R. Schrieffer, and A. J. Heeger, \textit{Solitons
in polyacetylene}, Phys. Rev. Lett. \textbf{42}, 1698 (1979).

\bibitem{Benalcazar17Science}W. A. Benalcazar, B. A. Bernevig, and
T. L. Hughes, \textit{Quantized electric multipole insulators}, Science
\textbf{357}, 61 (2017).

\bibitem{BBH17prb}W. A. Benalcazar, B. A. Bernevig, and T. L. Hughes,
\textit{Electric multipole moments, topological multipole moment pumping,
and chiral hinge states in crystalline insulators}, Phys. Rev. B \textbf{96},
245115 (2017).

\bibitem{Trifunovic19prx}L. Trifunovic and P. W. Brouwer, \textit{Higher-order
bulk-boundary correspondence for topological crystalline phases},
Phys. Rev. X \textbf{9}, 011012 (2019).

\bibitem{Schindler18SA}F. Schindler, A. M. Cook, M. G. Vergniory,
Z. Wang, S. S. P. Parkin, B. A. Bernevig, and T. Neupert, \textit{Higher-order
topological insulators}, Science Advances \textbf{4} 6 (2018).

\bibitem{Langbehn17prl}J. Langbehn, Y. Peng, L. Trifunovic, F. von
Oppen, and P. W. Brouwer, \textit{Reflection-symmetric second-order
oopological insulators and superconductors}, Phys. Rev. Lett. \textbf{119},
246401 (2017).

\bibitem{Khalaf18prb}E. Khalaf, \textit{Higher-order topological
insulators and superconductors protected by inversion symmetry}, Phys.
Rev. B \textbf{97}, 205136 (2018).

\bibitem{SongZD19prl}Z. Song, Z. Fang, and C. Fang, \textit{(d-2)
-dimensional edge states of rotation symmetry protected topological
states}, Phys. Rev. Lett. \textbf{119}, 246402 (2017).

\bibitem{Geier18prb}M. Geier, L. Trifunovic, M. Hoskam, and P. W.
Brouwer, \textit{Second-order topological insulators and superconductors
with an order-two crystalline symmetry}, Phys. Rev. B \textbf{97},
205135 (2018).

\bibitem{petrides2019arxiv}I. Petrides and O. Zilberberg, (2019),
arXiv:1911.08461 {[}cond-mat.mes-hall{]}.

\bibitem{WangZJ19prl}Z. Wang, B. J. Wieder, J. Li, B. Yan, and B.
A. Bernevig, \textit{Higher-order topology, monopole nodal lines,
and the origin of large Fermi arcs in transition metal dichalcogenides}
$XTe_{2}$ ($X=Mo,W$), Phys. Rev. Lett. \textbf{123}, 186401 (2019).

\bibitem{Schindler18NP}F. Schindler, Z. Wang, M. G. Vergniory, A.
M. Cook, A. Murani, S. Sengupta, A. Y. Kasumov, R. Deblock, S. Jeon,
I. Drozdov, H. Bouchiat, S. Gueon, A. Yazdani, B. A. Bernevig, and
T. Neupert, \textit{Higher-order topology in bismuth}, Nat. Phys.
\textbf{14}, 918 (2018).

\bibitem{Serra-Garcia18nature}M. Serra-Garcia, V. Peri, R. Sustrunk,
O. R. Bilal, T. Larsen, L. G. Villanueva, and S. D. Huber, \textit{Observation
of a phononic quadrupole topological insulator}, Nature \textbf{555},
342 (2018).

\bibitem{Peterson18nature}C. W. Peterson, W. A. Benalcazar, T. L.
Hughes, and G. Bahl, \textit{A quantized microwave quadrupole insulator
with topologically protected corner states}, Nature \textbf{555},
346 (2018).

\bibitem{Franca18prb}S. Franca, J. van den Brink, and I. C. Fulga,
\textit{An anomalous higher-order topological insulator}, Phys. Rev.
B \textbf{98}, 201114(R) (2018).

\bibitem{Thomale18np}S. Imhof, C. Berger, F. Bayer,
J. Brehm, L. W. Molenkamp, T. Kiessling, F. Schindler,
C.-H. Lee, M. Greiter, T. Neupert and R. Thomale, \textit{Topolectrical-circuit
realization of topological corner modes}, Nat. Phys. \textbf{14},
925 (2018).

\bibitem{YanZB18prl}Z. Yan, F. Song, and Z. Wang, \textit{Majorana
corner modes in a high-temperature platform}, Phys. Rev. Lett. \textbf{121},
096803 (2018).

\bibitem{WangQ18prl}Q. Wang, C.-C. Liu, Y.-M. Lu, and F. Zhang,\textit{
High-temperature Majorana corner states}, Phys. Rev. Lett. \textbf{121},
186801 (2018).

\bibitem{Hsu18prl}C.-H. Hsu, P. Stano, J. Klinovaja, and D. Loss,
\textit{Majorana Kramers pairs in higher-order topological insulators},
Phys. Rev. Lett. \textbf{121}, 196801 (2018).

\bibitem{Volpez19prl}Y. Volpez, D. Loss, and J. Klinovaja, \textit{Second-order
topological superconductivity in \textgreek{p} -junction Rashba layers},
Phys. Rev. Lett. \textbf{122}, 126402 (2019).

\bibitem{LiuT18prb}T. Liu, J. J. He, and F. Nori, \textit{Majorana
corner states in a two-dimensional magnetic topological insulator
on a high-temperature superconductor}, Phys. Rev. B \textbf{98}, 245413
(2018).

\bibitem{ZhuXY18prb}X. Zhu, \textit{Tunable Majorana corner states
in a two-dimensional second-order topological superconductor induced
by magnetic fields}, Phys. Rev. B \textbf{97}, 205134 (2018).

\bibitem{Shapourian18prb}H. Shapourian, Y. Wang, and S. Ryu, \textit{Topological
crystalline superconductivity and second-order topological superconductivity
in nodal-loop materials}, Phys. Rev. B \textbf{97}, 094508 (2018).

\bibitem{Ezawa18prl} M. Ezawa, \textit{Higher-order topological insulators
and semimetals on the breathing kagome and pyrochlore lattices}, Phys.
Rev. Lett. \textbf{120}, 026801 (2018).

\bibitem{okugawa2019arxiv}R. Okugawa, S. Hayashi, and T. Nakanishi,
(2019), arXiv:1907.01153 {[}cond-mat.mes-hall{]}.

\bibitem{wieder2019arxiv}B. J. Wieder, Z. Wang, J. Cano, X. Dai,
L. M. Schoop, B. Bradlyn, and B. A. Bernevig, (2019), arXiv:1908.00016
{[}cond-mat.mes-hall{]}.

\bibitem{FuL11prl}L. Fu, \textit{Topological crystalline insulators},
Phys. Rev. Lett. \textbf{106}, 106802 (2011).

\bibitem{Neupert18springer}T. Neupert and F. Schindler, ``Topological
crystalline insulators,'' in Topological Matter: Lectures from the
Topological Matter School 2017, edited by D. Bercioux, J. Cayssol,
M. G. Vergniory, and M. Reyes Calvo (Springer International Publishing,
Cham, 2018) pp. 31-- 61.

\bibitem{Jan17prx}J. Kruthoff, J. de Boer, J. van Wezel,
C. L. Kane, and R.-J. Slager, \textit{Topological classification
of crystalline insulators through band structure combinatorics}, Phys.
Rev. X \textbf{7}, 041069 (2017)

\bibitem{Yoshida18prb}A. Yoshida, Y. Otaki, R. Otaki, and T. Fukui,
\textit{Edge states, corner states, and flat bands in a two-dimensional
PT -symmetric system}, Phys. Rev. B \textbf{100}, 125125 (2019).

\bibitem{Khalaf18prx}E. Khalaf, H. C. Po, A. Vishwanath, and H. Watanabe,
\textit{Symmetry indicators and anomalous surface states of topological
crystalline insulators}, Phys. Rev. X \textbf{8}, 031070 (2018).

\bibitem{YuR11prb}R. Yu, X. L. Qi, A. Bernevig, Z. Fang, and X. Dai,
\textit{Equivalent expression of $Z_{2}$ topological invariant for
band insulators using the non-Abelian Berry connection}, Phys. Rev.
B \textbf{84}, 075119 (2011).

\bibitem{Fidkowski11prl}L. Fidkowski, T. S. Jackson, and I. Klich,
\textit{Model characterization of gapless edge modes of topological
insulators using intermediate Brillouin-zone functions}, Phys. Rev.
Lett. \textbf{107}, 036601 (2011).

\bibitem{Alexandradinata14prb}A. Alexandradinata, X. Dai, and B.
A. Bernevig, \textit{Wilson-loop characterization of inversion-symmetric
topological insulators}, Phys. Rev. B \textbf{89}, 155114 (2014).

\bibitem{khalaf19arxiv}E. Khalaf, W. A. Benalcazar, T. L. Hughes,
and R. Queiroz, (2019), arXiv:1908.00011 {[}cond-mat.meshall{]}.

\bibitem{Yang19arxiv}Y.-B. Yang, K. Li, L. M. Duan, and Y. Xu, (2019),
arXiv:1910.04151 {[}cond-mat.mes-hall{]}.

\bibitem{Yang2019arxiv}Y.-B. Yang, K. Li, and Y. Xu, (2019), arXiv:1910.14189
{[}cond-mat.mes-hall{]}.

\bibitem{Schnyder08prb}A. P. Schnyder, S. Ryu, A. Furusaki, and A.
W. W. Ludwig, \textit{Classification of topological insulators and
superconductors in three spatial dimensions}, Phys. Rev. B \textbf{78},
195125 (2008).

\bibitem{ChiuCK16rmp}C.-K. Chiu, J. C. Y. Teo, A. P. Schnyder, and
S. Ryu, \textit{Classification of topological quantum matter with
symmetries}, Rev. Mod. Phys. \textbf{88}, 035005 (2016).

\bibitem{chen17arxiv}B.-H. Chen and D.-W. Chiou, (2017), arXiv:1705.06913
{[}cond-mat.mes-hall{]}.

\bibitem{LiLH14prb}L. Li, Z. Xu, and S. Chen, \textit{Topological
phases of generalized Su-Schrieffer-Heeger models}, Phys. Rev. B \textbf{89},
085111 (2014).

\bibitem{Prezgonzlez18arxiv}B. Perez-Gonzalez, M. Bello, A. Gomez-Leon,
and G. Platero, (2018), arXiv:1802.03973 {[}cond-mat.meshall{]}.

\bibitem{Kang19prb}B. Kang, K. Shiozaki, and G. Y. Cho, \textit{Many-body
order parameters for multipoles in solids}, Phys. Rev. B \textbf{100},
245134 (2019).

\bibitem{Wheeler19prb}W. A. Wheeler, L. K. Wagner, and T. L. Hughes,
\textit{Many-body electric multipole operators in extended systems},
Phys. Rev. B \textbf{100}, 245135 (2019).

\bibitem{roy19arxiv}B. Roy, (2019), arXiv:1906.10685 {[}cond-mat.mes-hall{]}.

\bibitem{Ono19prb}S. Ono, L. Trifunovic, and H. Watanabe, \textit{Difficulties
in operator-based formulation of the bulk quadrupole moment}, Phys.
Rev. B \textbf{100}, 245133 (2019).

\bibitem{LiC18prb}C.-A. Li, S.-B. Zhang, and S.-Q. Shen, \textit{Hidden
edge Dirac point and robust quantum edge transport in InAs/GaSb quantum
wells}, Phys. Rev. B 97, 045420 (2018).

\bibitem{Skolasinski18prb}R. Skolasinski, D. I. Pikulin, J. Alicea,
and M. Wimmer, \textit{Robust helical edge transport in quantum spin
Hall quantum wells}, Phys. Rev. B \textbf{98}, 201404(R) (2018).

\bibitem{Haldane88prl}F. D. M. Haldane, \textit{Model for a quantum
Hall effect without Landau levels: condensed-matter realization of
the \textquotedbl parity anomaly\textquotedbl}, Phys. Rev. Lett.
\textbf{61}, 2015 (1988).

\bibitem{zhang19arxiv}S.-B. Zhang and B. Trauzettel, \textit{Detection
of second-order topological superconductors by Josephson junctions},
Phys. Rev. Research \textbf{2}, 012018(R) (2020).

\bibitem{You2019arXiv}Y. You, (2019), arXiv:1908.04299 {[}cond-mat.str-el{]}.

\bibitem{LiJian09prl}J. Li, R.-L. Chu, J. K. Jain, and S.-Q. Shen,
\textit{Topological Anderson insulator}, Phys. Rev. Lett. \textbf{102},
136806 (2009).

\bibitem{Shem14prl}I. Mondragon-Shem, T. L. Hughes, J. Song, and
E. Prodan, \textit{Topological criticality in the chiral-symmetric
AIII class at strong disorder}, Phys. Rev. Lett. \textbf{113}, 046802
(2014).

\bibitem{LiCA19}C.-A. Li, \textit{et al}., unpublished.
\end{thebibliography}
\end{document}